\documentclass[%
 reprint,
 superscriptaddress,
 amsmath,amssymb,
 aps,
 longbibliography
]{revtex4-2}

\usepackage{graphicx}
\usepackage{dcolumn}
\usepackage{bm}
\usepackage[separate-uncertainty=true]{siunitx}

\renewcommand{\vec}[1]{\mathbf{#1}}

\usepackage{xcolor}

\begin{document}


\title{Experimental Demonstration of Coupled Learning in Elastic Networks}

\author{Lauren E. Altman}
\email{laltman2@sas.upenn.edu}
\affiliation{Department of Physics \& Astronomy, University of Pennsylvania, Philadelphia, PA 19104, USA}

\author{Menachem Stern}
\affiliation{Department of Physics \& Astronomy, University of Pennsylvania, Philadelphia, PA 19104, USA}

\author{Andrea J. Liu}
\affiliation{Department of Physics \& Astronomy, University of Pennsylvania, Philadelphia, PA 19104, USA}

\author{Douglas J. Durian}
\affiliation{Department of Physics \& Astronomy, University of Pennsylvania, Philadelphia, PA 19104, USA}
\affiliation{Department of Mechanical Engineering and Applied Mechanics, University of Pennsylvania, Philadelphia, PA 19104, USA}

\date{\today}

\begin{abstract}
Coupled learning is a contrastive local learning scheme for 
tuning the properties of individual elements 
within a network to achieve 
desired functionality of the system. 
It takes advantage of physics both 
to learn using local rules and 
to ``compute” the output response to input data, 
thus enabling the system to perform 
decentralized computation without the need for 
a processor or external memory. 
We present three proof-of-concept mechanical networks of increasing complexity, and demonstrate how they can learn tasks such as self-symmetrization and node allostery
via iterative tuning of individual spring rest lengths. 
These mechanical networks could feasibly be scaled 
and automated to solve increasingly complex tasks, 
hinting at a new class of smart metamaterials.
\end{abstract}

\maketitle

\section{Introduction}

Learning is a physical process by which a system evolves to exhibit a desired behavior. 
Even algorithms such as artificial neural networks, which are trained via global optimization algorithms such as gradient descent \cite{lecun_deep_2015}, are carried out in digital hardware, incurring a significant energy cost.

Another approach to learning uses computational global optimization (\textit{e.g.} by backpropagation~\cite{Rojas1996}) to learn or train but
``computes," or calculates outputs in response to inputs (performs inference) in hardware by applying boundary conditions as inputs and using physics to propagate information through the system to ``compute" an output response. Such a hybrid approach was used to produce laboratory materials with a desired localized strain response to a localized strain applied elsewhere~\cite{rocks_designing_2017,reid2018auxetic}, following earlier ideas~\cite{goodrich_principle_2015}. Pre-trained artificial neural networks have been implemented in physical systems \cite{wright_deep_2022} to perform more complex tasks. Most recently,
Lee \textit{et al.} \cite{lee2022mechanical} used computation to train a network of elastic components with tunable stiffnesses using voice coils.  
However, hybrid approaches can fail when the model used to compute updates does not approximate the real system well enough. Even for an excellent model, training on the computer becomes more expensive and less accurate as the system increases in size. 
As a result, there is a substantial advantage to using approaches that avoid computation altogether during training.

Computational global optimization uses global information about the state of the entire network to compute gradients, in order to determine how the system should evolve in the next training step. 
Biological learning systems, as well as physical learning systems that eschew processors, do not have access to such global information as they evolve, and so must instead rely on decentralized learning schemes based on local update rules to achieve desired behaviors \cite{stern2023learning}. 
In such systems, learning is better described as an emergent phenomenon in a complex material. 
Decentralized local learning has multiple advantages over the computational global optimization approach \cite{deneve_brain_2017}, including  power efficiency \cite{sengupta_power_2014,stern2024training}, scalability of processing speed with network size \cite{john2020self}, and robustness to damage \cite{mcgovern_hemispherectomy_2019,dillavou2022demonstration}.
In this approach, training occurs entirely in-situ on a local, element-wise level. 

Specifically, consider physical networks of nodes connected by edges, in which edges update in a distributed manner based on their own responses and properties, so that the time required to compute an update does not significantly increase as the number of edges increases.
The poor scaling and computational expense of traditional neural networks that use global rules is one of the major bottlenecks to progress in the field of artificial learning, and raises concerns about energy consumption \cite{de2023growing}.
Networks that learn ``on-the-fly" via local training schemes also benefit from robustness to damage: if part of the network is lost, the distributed elements can learn a new solution to the desired task \cite{dillavou2022demonstration}.

Approaches such as directed aging \cite{pashine_directed_2019,hexner_effect_2020,hexner_periodic_2020}, contrastive learning~\cite{movellan_contrastive_1991,pashine_local_2021}, and nudged contrastive learning such as equilibrium propagation~\cite{scellier_equilibrium_2017, kendall_training_2020,martin_eqspike_2021} and coupled learning~\cite{stern_supervised_2021,dillavou2022demonstration,dillavou2023machine}, can be implemented in experiment alone, relying on the each edge's response to an applied boundary condition to tune properties of individual edges. Pashine, \textit{et al.} used directed aging to produce laboratory networks with negative Poisson's ratios~\cite{pashine_directed_2019, hexner_effect_2020}, and Arinze, \textit{et al.} used local rules akin to directed aging to train thin elastic sheets to fold in desired ways in response to force patterns \cite{arinze2023learning,stern_supervised_2020}. Directed aging, however,
is limited in its use cases because it does not minimize the true cost function of the training task unless the desired response is close to the untrained response. 

Approaches based on contrastive learning~\cite{movellan_contrastive_1991,pashine_local_2021} compare a network's internal state in response to different sets of boundary conditions. In particular, nudged contrastive local learning approaches either reduce to gradient descent in the limit of infinitesimal nudges~\cite{scellier_equilibrium_2017} or have large enough projections onto the direction of gradient descent~\cite{stern_supervised_2021} to solve complex tasks effectively~\cite{kendall_training_2020,martin_eqspike_2021,dillavou2023machine,scellier2023energy}.  A variety of approaches for implementing these ideas in hardware have been advanced~\cite{martin_eqspike_2021, AnisettiPRR2023,falk2023learning}. Coupled learning has been demonstrated in the lab~\cite{dillavou2022demonstration, wycoff_desynchronous_2022, stern2022physical} to learn various tasks, including classification, but so far only Pashine~\cite{pashine_local_2021} has used contrastive learning to obtain systems with a desired mechanical behavior (negative Poisson's ratio).  

A mechanical network with tunable mechanical edge properties that implements nudged contrastive learning poses several challenges for laboratory construction.
The physical degrees of freedom in a mechanical network, namely the node positions, are vectorial in nature, rather than scalar like the voltages at nodes of an electrical network. This increases the dimensional complexity of the system.
A mechanical network is also spatially embedded, so the architecture and connectivity of the network must be carefully chosen for proper network coordination and to avoid collision of neighboring edges as well as mechanical instabilities such as frustration or buckling. It is also a nontrivial engineering problem to develop mechanical elements with adjustable learning degrees of freedom.

A recent work introduces directed springs~\cite{patil2023self} whose stiffness evolves in response to its physical behavior. 
Here we introduce a different approach, in which we tune each edge's \emph{rest length}, rather than its \emph{stiffness}. In principle, either set of learning degrees of freedom can be effective in learning; Refs.~\cite{pashine_directed_2019,hexner_effect_2020,hexner_periodic_2020} refer to models with tunable stiffness as $k$-models and those with tunable rest lengths as $\ell$-models. Note that stiffness is directly analogous to conductance in electrical networks, while rest length has no direct analogy and is unique to mechanical systems, so is interesting to study in its own right. 

In our approach, a turnbuckle is connected to a Hookean spring at its end. By adjusting the length of the turnbuckle, the effective rest length of combination spring-turnbuckle edge is altered. 
Since updates are not automated but rather made manually in our prototype, it is imperative to choose architectures that minimize the number of tunable edges while still exhibiting a diversity of tunable behaviors.  
Here we provide three laboratory demonstrations of experimental mechanical networks that can be trained using coupled learning with adjustable rest length components, and we compare with simulation.
We show that the three architectures effectively span the range of possible states to achieve the desired behavior quickly, smoothly, and effectively. 
Our experiments demonstrate how real-world considerations like friction and buckling cause deviations from simulated learning behavior, and highlight unexpected difficulties of applying coupled learning to mechanical systems.
Despite these limitations, our initial successes show promise for learning more complex functionalities in larger networks, and can serve as a guide for future implementations of elastic learning systems.

\section{Coupled Learning for Spring-Turnbuckle Systems}

\begin{figure*}
    \centering
    \includegraphics[width=0.8\textwidth]{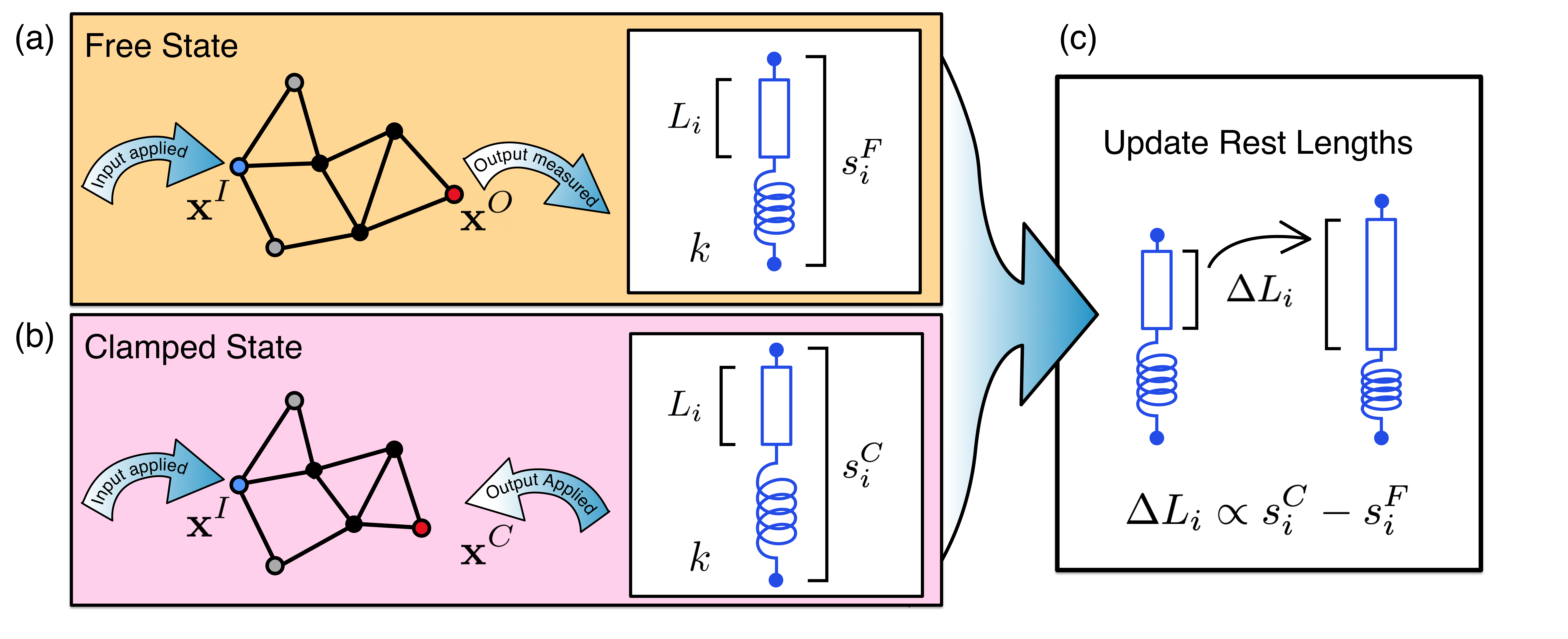}
    \caption{Schematic detailing the coupled learning algorithm for an arbitrary elastic network. 
    A mechanical network is constructed such that each $i^\text{th}$ edge consists of a spring with stiffness $k$ connected to a turnbuckle with adjustable rest length $L_i$.
    The springs act as the edges of the network, and their connection points are nodes. 
    Specific nodes are chosen as ``inputs" and ``outputs" for the network to 
    learn a desired task.
    Some nodes (gray) are fixed to prevent translation and rotation of the entire network, and the remaining nodes move in response to imposed boundary conditions.
    (a) In the free state, the position of the input node $\vec{x}^I$ is enforced, and the position of the output node is measured $\vec{x}^O$. Each $i^\text{th}$ edge has length (\textit{i.e.} node-node separation) $s_i^F$. 
    (b) In the clamped state, the input node's position is still fixed at $\vec{x}^I$ and the output node is ``clamped" to position $\vec{x}^C$. Each $i^\text{th}$ has length $s_i^C$.
    (c) By locally comparing the lengths of each edge, the update rule in Eq.~\eqref{eq:discretelearningrule} determines how $L_i$ evolves.}
    \label{fig:coupled_learning}
\end{figure*}

We construct a mechanical network such as the one depicted in Fig.~\ref{fig:coupled_learning}(a).
Each edge of this network is an extension spring with the same stiffness $k$ and rest length $l$ connected to a rigid turnbuckle.  The turnbuckle for spring $i$ allows it to have an adjustable length $L_i$, as drawn schematically in the inset of Fig.~\ref{fig:coupled_learning}(a). 
The total node-node separation of the edge, including the length of the rigid part, is $s_i$, such that the spring's extension past its rest length is $s_i - (l + L_i)$. 
The mechanical energy associated with this edge is given by Hooke's law as
\begin{equation}
\label{eq:energy}
    u_i = \frac{1}{2} k \left[ s_i - \left(l + L_i\right)\right]^2.
\end{equation}
The total mechanical energy in the network is the sum of the elastic energies of all edges,
\begin{equation}
\label{eq:totalenergy}
    U = \sum_j u_j,
\end{equation}
and serves as the ``physical cost function" which the system automatically minimizes with respect to the node positions, namely the ``physical degrees of freedom," subject to the imposed boundary conditions. 

We wish to train this network to achieve some desired behavior in the positions of its nodes.
This is what is referred to as a ``motion task."
We select certain nodes to be ``inputs" and ``outputs" for our task, depicted in Fig.~\ref{fig:coupled_learning}(a) as blue and red filled circles, respectively. 
When the input node is at position $\vec{x}^I$, we want the output node's position $\vec{x}^O$ to be at some desired value, $\vec{x}^D$. 

We train for this behavior by iteratively adjusting the turnbuckle lengths at each edge according to the coupled learning algorithm \cite{stern_supervised_2021}. 
Here, the node positions, or physical degrees of freedom, equilibrate so that there is no net force on each edge, while the rest lengths act as ``learning degrees of freedom" to be adjusted during training. 
In the free state, Fig.~\ref{fig:coupled_learning}(a), we apply the input boundary condition to the network by fixing the position of the input node, $\vec{x}^I$, and allowing all other physical degrees of freedom to equilibrate.
The output node's position $\vec{x}^O$ is measured, along with the extensions of each edge in the network, $s_i^F$.
In the clamped state, Fig.~\ref{fig:coupled_learning}(b), we once again apply the input boundary condition, but now we also enforce a fixed value for the position of the output node by ``nudging" it towards the desired position $\vec{x}^D$. 
The clamped output value $\vec{x}^C$ generally takes the form 
\begin{equation}
\label{eq:clamp}
    \vec{x}^C = \vec{x}^F + \eta (\vec{x}^D - \vec{x}^F ) ,
\end{equation}
where the ``nudge factor" $0 < \eta \leq 1$ is a hyperparameter of the training scheme. In all experiments below, we take a full nudge of $\eta=1$ so the output node is clamped at the desired position at each learning step. The lengths $s_i^C$ of each edge in this state are also recorded.

In coupled learning, the system evolves by comparing the mechanical energy of the network in the free and clamped states.
Learning is achieved when the energy in the clamped state $U^C$ is equal to the energy in the free state $U^F$.
In the absence of nonlinear mechanical effects like buckling, 
the difference between these energies, which we refer to as the ``learning contrast function", is always non-negative because the clamped state is more strongly constrained than the free state, $U^C - U^F  \geq 0$.
Analogous to the machine-learning approach of minimizing a loss function like mean-squared error (MSE), the rest lengths evolve by descending along the gradient of the contrast function:
\begin{eqnarray}
    \frac{dL_i}{dt} &\propto& -\frac{\partial}{\partial L_i}\left[\sum u_j^C-u_j^F \right] \label{eq:coupledlearning} \\
        &=& -\frac{\partial}{\partial L_i}(u_i^C-u_i^F) \label{eq:purelylocal} \\
        &=& k(s_i^C-s_i^F) \label{eq:localsimplified}
\end{eqnarray}
Since the learning contrast function was not squared, the partial derivative in Eq.~(\ref{eq:coupledlearning}) picks out only the $j=i$ term, which is readily simplified to Eq.~(\ref{eq:localsimplified}) using Eq.~(\ref{eq:energy}). Thus, we arrive at the following general discrete learning rule for each edge's update over a learning step for any network of identical spring-turnbuckle edges:
\begin{equation}
    \Delta L_i = \alpha(s_i^C-s_i^F)
\label{eq:discretelearningrule}
\end{equation}
where $\alpha$ is a per-step learning rate that we shall set in experiment, and $(s_i^C-s_i^F)$ is the difference in clamped and free lengths of the edge being updated.  This simple rule is purely local: each edge is updates according only to its behavior, irrespective of how other edges change upon clamping. In the experiments presented below, we train with different learning rate parameters in the range $0.1<\alpha\le1$. Iterative updates should drive the global learning contrast function to zero in order to achieve the desired motion function. 
Appendix~\ref{sec:alignment} details the conditions under which the local and global update rules align.

While our Eq.~(\ref{eq:discretelearningrule}) update rule is spatially local, it is not \emph{temporally} local because the learning rule requires simultaneous information about the system in two states. 
The experimental implementation in an 
electrical resistor network \cite{dillavou2022demonstration} was able to circumvent this issue 
by building identical twin networks to run the 
free and clamped states simultaneously.
By contrast, the mechanical network is embedded in space, 
posing difficulties for constructing twin 2-dimensional networks 
side-by-side, and an impossibility entirely for 3-dimensional implementations.
Therefore, our approach must rely on temporal memory of the spring extensions between the free and clamped states.

\section{Motion Divider}
\label{sec:motiondivider}

Analogous to the voltage divider in the electrical network case, our first demonstration is a ``motion divider" network, or two tunable-rest length springs connected in series.
A photograph of the network is shown in Fig.~\ref{fig:motion_divider}(a). 
We construct the network so it is hanging vertically under gravity from a fixed point, thereby restricting the motion to be along the axis of gravity.
We define the one-dimensional position $y=0$ to be at the fixed hanging point, and $y>0$ measures position downwards from this origin.
The first spring is connected to the $y=0$ fixed point via a turnbuckle of length $L_1$, terminating at position $y_1$. 
The second turnbuckle of length $L_2$ then connects to the second spring, which terminates at position $y_2$.
We choose $y_2$ to be the input node of our system and $y_1$ to be the output node. 
Both springs have equal stiffnesses $k$ and natural rest lengths $l$, so that the Eq.~(\ref{eq:discretelearningrule}) learning rule applies.
The forces acting on springs in series are equal, so $k (y_1 - l - L_1) = k (y_2 - y_1 - l - L_2)$ holds and can be solved for the output node position as
\begin{equation}
    \label{eq:motiondividerEOM}
    y_1 = \frac{1}{2}y_2 + a
\end{equation}
where $a = \frac{1}{2}(L_1 - L_2)$. 
Choosing some desired fixed value for $a$ defines a trainable task for our system. 
This is equivalent to a machine-learning linear regression problem with one variable coefficient \cite{HOPE202067}.
Note that there is no unique solution for $L_1$ and $L_2$; only their difference is trained for. Like typical machine learning algorithms, our system is over-parameterized and thus has multiple solutions.

\subsection{Apparatus}
Each unit cell of our network consists of
an extension spring coupled to a turnbuckle in series, as
shown diagrammatically in Fig.~\ref{fig:coupled_learning} and 
photographically in Fig.~\ref{fig:motion_divider}(a).
The spring has a stiffness of $k = \SI{13}{N/cm}$ 
and a rest length of $l = \SI{5.5}{cm}$ (Grainger 5108N536), while the turnbuckle has a 
range of lengths between $L^{min} = \SI{12}{cm}$ and $L^{max} = \SI{16}{cm}$ (eoocvt M4 Stainless Steel 304). 
Updates are made on the system via turns of the turnbuckle, 
where each half-turn results in a change in length of $\Delta L /\text{turn} = \SI{0.079}{cm}$.
The total effective rest length of the unit cell object is therefore 
$l + L_i$.
These unit cells are connected together using keyrings, which 
serve functionally as the nodes of our network. 
The system is clamped at the nodes manually by inserting a small rod through the keyrings and fixing its position up or down using a clamp on a vertical pole. 
Measurements were made manually by the experimenter using a ruler.

\subsection{Results}

\begin{figure}
    \centering
    \includegraphics[width=0.9\columnwidth]{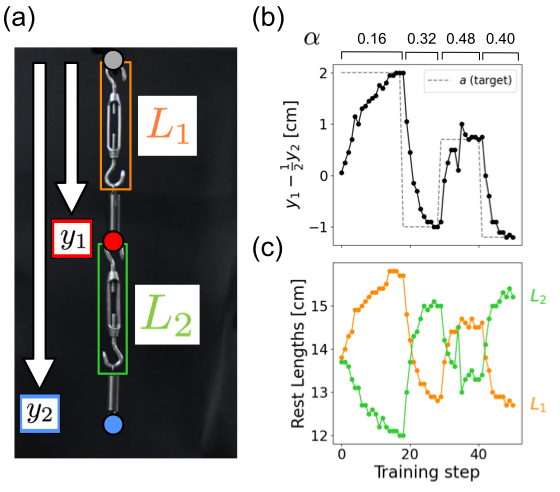}
    \caption{Training a mechanical motion divider to obtain desired behavior in the form of an additive constant $a$. 
    (a) Photograph of the experimental apparatus. Two springs in series are hung vertically from a fixed point.
    The positions of the two nodes relative to the fixed point, $y_1$ and $y_2$, serve as the target and source for the training task, respectively. The learning degrees of freedom, $L_1, L_2$, determine the relationship between the node positions.
    (b) Results of training the network four consecutive times. Different goal values for $a$ (dashed gray line) and learning rates $\alpha$ were used for each consecutive training.
    (c) Evolution of the learning degrees of freedom over the course of training. $L_1$ (orange) evolves inversely to $L_2$ (green) as the network approaches the goal state. Rest lengths were not reset to their initial values after each successful training.}
    \label{fig:motion_divider}
\end{figure}

Fig.~\ref{fig:motion_divider}(b) shows the evolution of the node couplings over the course of training. 
The network is trained multiple times to achieve different values for the goal state, $a$. 
The turnbuckle lengths were not reset after each 
training; instead, the network was able to re-learn a new task without initialization. 
The evolution of the learning degrees of freedom, $L_1, L_2$ is shown in Fig.~\ref{fig:motion_divider}(c). 
Each time we change the training task $a$, we also choose a different learning rate to demonstrate that the network may be trained at different time scales.

Training time required to reach the desired state is determined both by the learning rate and the distance in parameter space between the network's initial and final states.
The physical and learning degrees of freedom are expected to evolve exponentially and asymptotically approach their desired values over the course of training. 
For a full derivation of the learning dynamics, see Appendix~\ref{sec:dynamics}.
Fig.~\ref{fig:MD_evolution} presents each of the four trials of training separately. 
Each of these is fit to the derived time-evolution Eq.~\eqref{eq:md_time_evolution}, where $\nu t$ has been replaced by $\alpha n$, with $n$ being the number of training steps.
The fits are overlaid on the data in red.
The table below compares the true values for $a$ and $\alpha$ with those obtained from the exponential fit, and finds good agreement, with an average error of \SI{12}{\%}.
By comparing fitted values for $a$ with the true values, we may also obtain an estimate of the measurement error, where $\epsilon \sim \langle \Delta a \rangle = \SI{0.1}{cm}$.

\begin{figure}
    \centering \includegraphics[width=0.45\textwidth]{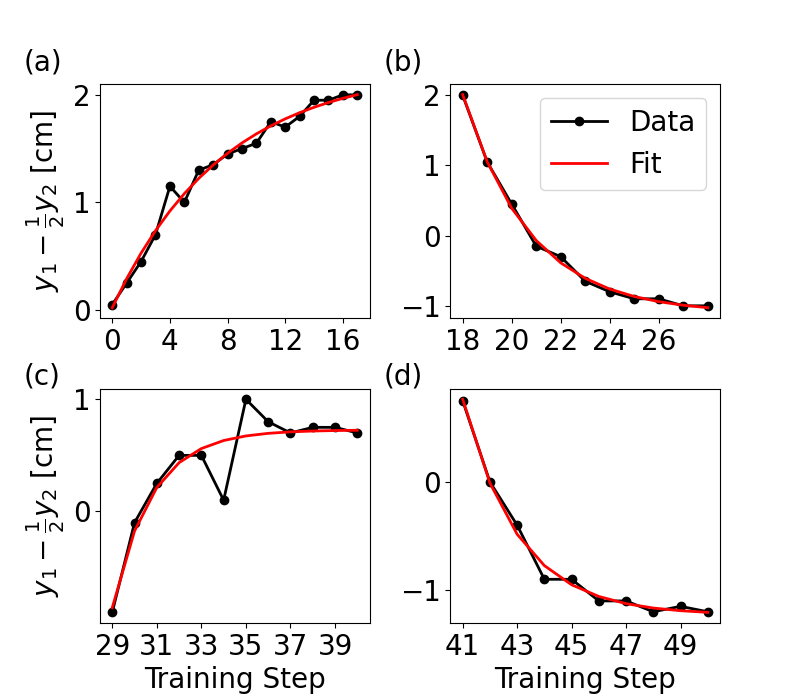}
    \vspace{1em}
      \begin{ruledtabular}
    \centering
    \begin{tabular}{ccccc}
    Trial & True $a$ [cm] & Fit $a$ [cm] & True $\alpha$ & Fit $\alpha$\\
    \colrule
    (a) & 2.00 & 2.25 & 0.16 & 0.13 \\
    (b) & -1.00 & -1.11 & 0.32 & 0.37 \\
    (c) & 0.70 & 0.73 & 0.48 & 0.56 \\
    (d) & -1.20 & -1.23 & 0.40 & 0.49 \\
    \end{tabular}
    \end{ruledtabular}
    \caption{Comparison of motion divider learning dynamics with theoretical prediction. 
    The time-evolution of the physical behavior for each of the four trials from Fig.~\ref{fig:motion_divider}(b) are separately fit to an exponential form and values for the goal state $a$ and the learning rate $\alpha$ are obtained. 
    The table (bottom) compares these fitted values with the true experimental values for $a$ and $\alpha$ for each of the four trials.}
    \label{fig:MD_evolution}
\end{figure}

\section{Symmetry Network}
\label{sec:symmetrynet}

\begin{figure*}
    \centering
    \includegraphics[width=0.8\textwidth]{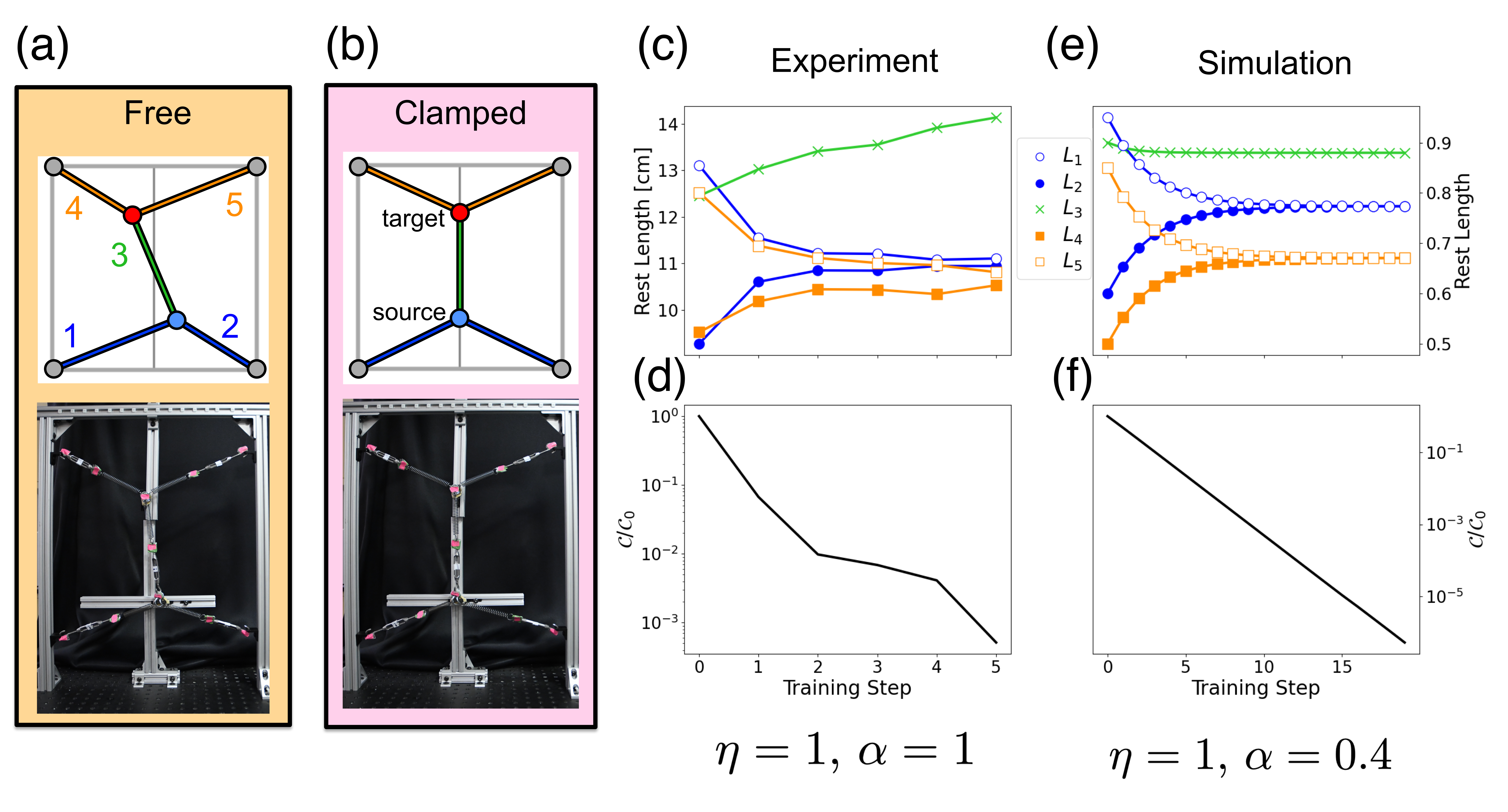}
    \caption{Training a two-dimensional network for symmetry.
    (a-b) Schematics and photographs of the free and clamped states of the network for a given initial configuration. Edges are labeled by number (a, top). Photographs of the experimental apparatus show marker tracking system and clamping mechanism.
    (c-d) Experimental training results. Rest lengths (c) converge to the symmetric state within experimental precision. 
    The cost $\mathcal{C} = U_C - U_F$ (d) decreases by three orders of magnitude. 
    (e-f) Simulation training results. The rest lengths (e) evolve such that $L_1 = L_2$ and $L_4 = L_5$. 
    Cost (f) decreases exponentially by five orders of magnitude. 
    }
    \label{fig:symmetry}
\end{figure*}

We further demonstrate our system's ability to learn more complex tasks by constructing a two-dimensional network within a frame. 
The network is shown schematically and photographically in Fig.~\ref{fig:symmetry}(a-b). 
Four of the edges are fixed to rigid a $\SI{45}{cm} \times \SI{40}{cm}$ frame constructed from 80/20, with the fifth central edge connecting the two internal nodes of the system. 
Edges are typically stretched past their rest length value by about $\SI{6}{cm}$ to prevent buckling.
Edges are labeled by number and nodes are denoted as source (blue) and target (red).

The goal is to train this network to become left-right symmetric,
which occurs when $L_1 = L_2$ and $L_4 = L_5$. This task may be encoded in the physical degrees of freedom (node positions) of the network with the condition: 
\begin{center}
\label{eq:symmetrytask}
    $x_{\text{target}} = x_{\text{source}} = 0$ for \textit{any} $y_{\text{source}}$ value,
\end{center}
where $x=0$ is defined to be the midline between the fixed nodes on either side.
This choice of task is not unique: other tasks can be ``solved" by the desired symmetric state, see Appendix~\ref{sec:onetarget}.
We may then train the network for this one-input, two-output task using a single data point. 

\subsection{Experiment}
\label{sec:symmetry_exp}
The symmetry network was constructed using the same unit cell design as in the motion divider in Section \ref{sec:motiondivider}. 
The springs used here have a natural rest length of $l = \SI{5.5}~{cm}$ and a stiffness of \SI{27.8}~{N/cm} (Grainger 1NAA2).
The external frame of the network was constructed from 80/20 parts and eyeholes screwed into the rail serve as the fixed nodes. 

Training this network requires only one input-output pair. 
Both outputs, $x_{\text{source}}$ and $x_{\text{target}}$, have desired values that are directly central in the frame.
Since training can occur for any choice of the input value, $y_{\text{source}}$, we choose the equilibrium position of this node in its initial state for simplicity. 

It is important to note that the physical degrees of freedom of a single node are mixed: the ``source" node acts as an input in the $y$-direction and as an output in the $x$-direction. 
The coupled learning algorithm allows for this seemingly strange coupling, and the network is still trainable under this scheme.
The mixing of degrees of freedom was achieved experimentally by aligning the nodes along 80/20 tracks, as can be seen in the photos in Fig.~\ref{fig:symmetry}(a-b). 
In the free state, we fix the input value for the training task, $y_{\text{source}}$, by attaching the node to a slide-in nylon tool hanger which slots into a horizontal 80/20 track so its vertical position is fixed.
The node may still move along the horizontal direction with a small coefficient of static friction $\mu \approx \num{0.1}$ for nylon on dry aluminum.
The target node is left unconstrained so both its degrees of freedom can freely equilibrate.
In the clamped state, we fix both degrees of freedom of the bottom node, $ \{ y_{\text{source}}, x_{\text{source}} \}$, by placing stoppers along the horizontal track described previously.
The horizontal position of the top node, $x_{\text{target}}$ is restricted by placing it on a vertical track using the same tool hanger as above, while its vertical position $y_{\text{target}}$ is allowed to equilibrate freely.

Since this network requires measurements in two dimensions, we automate the measurement process using a digital camera.
As can be seen in the photographs in Fig.~\ref{fig:symmetry}(a-b), 
we attach pink markers to all \num{6} nodes, as well as to the \num{5} points of connection between the turnbuckles and springs. 
The camera captures a three-channel color image of the entire network, where we choose a beneficial white balance (\SI{2500}{K}) and tint (M6.0) to maximize the contrast between the markers and the remainder of the image.
We then split the color channels and subtract the green channel from the red, leaving us with a one-channel image with high intensity values at the pixels associated with markers and low intensity everywhere else. 
The image is then binarized using a threshold intensity value of \num{100}, which is in between the well-separated high and low intensity values.
The pixel values that have been identified after binarization are then clustered into \num{11} clusters using a k-means algorithm \cite{lloyd1982least}.
The centroid of each cluster is determined to be the $(x,y)$ position of the associated marker.
This allows us to track all node positions, spring extensions, and turnbuckle lengths at every stage of the experiment.
For a given photograph with \num{11} markers such as the ones in Fig.~\ref{fig:symmetry}, the full processing and tracking takes approximately \SI{0.7}{s} on a standard Macbook Pro.

The results of training the physical network are shown in Fig.~\ref{fig:symmetry}(c-d). 
The learning degrees of freedom evolve such that $L_1$ comes to meet $L_2$ and $L_4$ meets $L_5$ within measurement error of \SI{0.28}~{cm} in only \num{5} training steps.
The time evolution of these rest lengths follows a similar trend as in the motion divider, where the largest adjustments happen in the early training steps and updates become asymptotically small as the network approaches the learned state. 
See Appendix~\ref{sec:dynamics} for more information about learning dynamics.
The success of training may also be measured by the cost function, $\mathcal{C} = U_C - U_F$.
Fig.~\ref{fig:symmetry}(d) shows that the cost decreases exponentially from its initial value by about three orders of magnitude overbthe course of training. 

The middle edge $L_3$ does not exhibit exponential time-evolution behavior, but rather drifts upward over the course of training by a total of \SI{1.5}{cm}, or \SI{37.5}{\%} of the range of values.
This linear evolution suggests some systematic experimental bias within the system, which may be a result of friction or mechanical instability.
Since $L_3$ has no bearing on the desired state, the network can learn the task for any value of $L_3$.

\subsection{Simulation}
There are many different ways to train for symmetry in our two-dimensional network. 
Our goal for this task is a specified internal state of the network, but the coupled learning scheme requires that we specify a \emph{behavior} in the nodes that is satisfied by this internal state, of which there are multiple options.
We therefore use simulation support to explore the different iterations of our task and examine their behavioral dynamics. 
We may also use simulation to modify the aspect ratio of the frame, as the angles of the edges at the internal nodes contributes to the coupling strength of the learning signal.
Having done this, we chose the training task and geometry that allowed for the most efficient evolution to the learned state for our experimental demonstration.

We simulated training on this network using the FIRE optimization algorithm to determine the network's node positions when boundary conditions are applied \cite{bitzek2006structural}.  
Here, each edge was allowed to vary between \num{0.5} and \num{1.0} length units, and all edge stiffnesses were fixed at the same value, \num{1}. 
Initial edge lengths were selected at random with validation that the network was sufficiently detuned from its goal state. 

The behavior that was used to train the network is a one-input / two-output task, but we could have alternatively defined a one-input, one-output task that is equally satisfied by the symmetric state of the network.
Training on the one-input, one-output task was successful in simulation, but required very long training times that were unfeasible for an experiment with manual operation. 
In this case, $L_4$ and $L_5$ evolve quickly to their desired values due to their strong coupling to the target node. 
$L_1$ and $L_2$, however, are only connected directly to the source node, which does not move in position very much between the free and clamped states, and therefore evolve very slowly. 
For more information, see Appendix~\ref{sec:onetarget}.
The two-output task allows for a strong learning signal for all of the edges in the network.

Training was performed using a full nudge $\eta = 1$ and Eq.~(\ref{eq:discretelearningrule}) with a learning rate of $\alpha = 0.4$. 
Successful training is possible with a higher learning rate, but we have artificially slowed the training to better examine the learning dynamics.
Fig.~\ref{fig:symmetry}(e) shows the evolution of the edges' rest lengths during the course of training. The outer edges evolve smoothly to meet their desired state, $L_1=L_2$ and $L_4 = L_5$, in approximately 10 training steps. 

The middle edge's rest length $L_3$ has no bearing on the desired goal state, yet it still evolves during training, particularly in the first few training steps. 
The desired state is not dependent on $L_3$, but the cost function is, as this is the edge which directly connects the two relevant nodes.
While evolution of $L_3$ is not required for the network to learn, allowing it to vary during training according to the coupled learning update rule helps the network descend down the cost landscape more quickly. 

In the experimental trial, Fig.~\ref{fig:symmetry}(c), $L_3$ systematically drifts upward, rather than the asymptotic decrease that is seen in Fig.~\ref{fig:symmetry}(e).
While this likely inhibited the speed of training, the network was still able to converge on the desired configuration of rest lengths.
This lends credence to the coupled learning algorithm's robustness even in non-ideal experimental conditions.
In our contrastive local scheme, training can fail to converge for individual edges on a local level, but the network as a whole can succeed in learning the desired behavior.

\section{Randomized Nonlinear Network}
\label{sec:nonlinear_net}

\begin{figure*}
    \centering
    \includegraphics[width=0.8\textwidth]{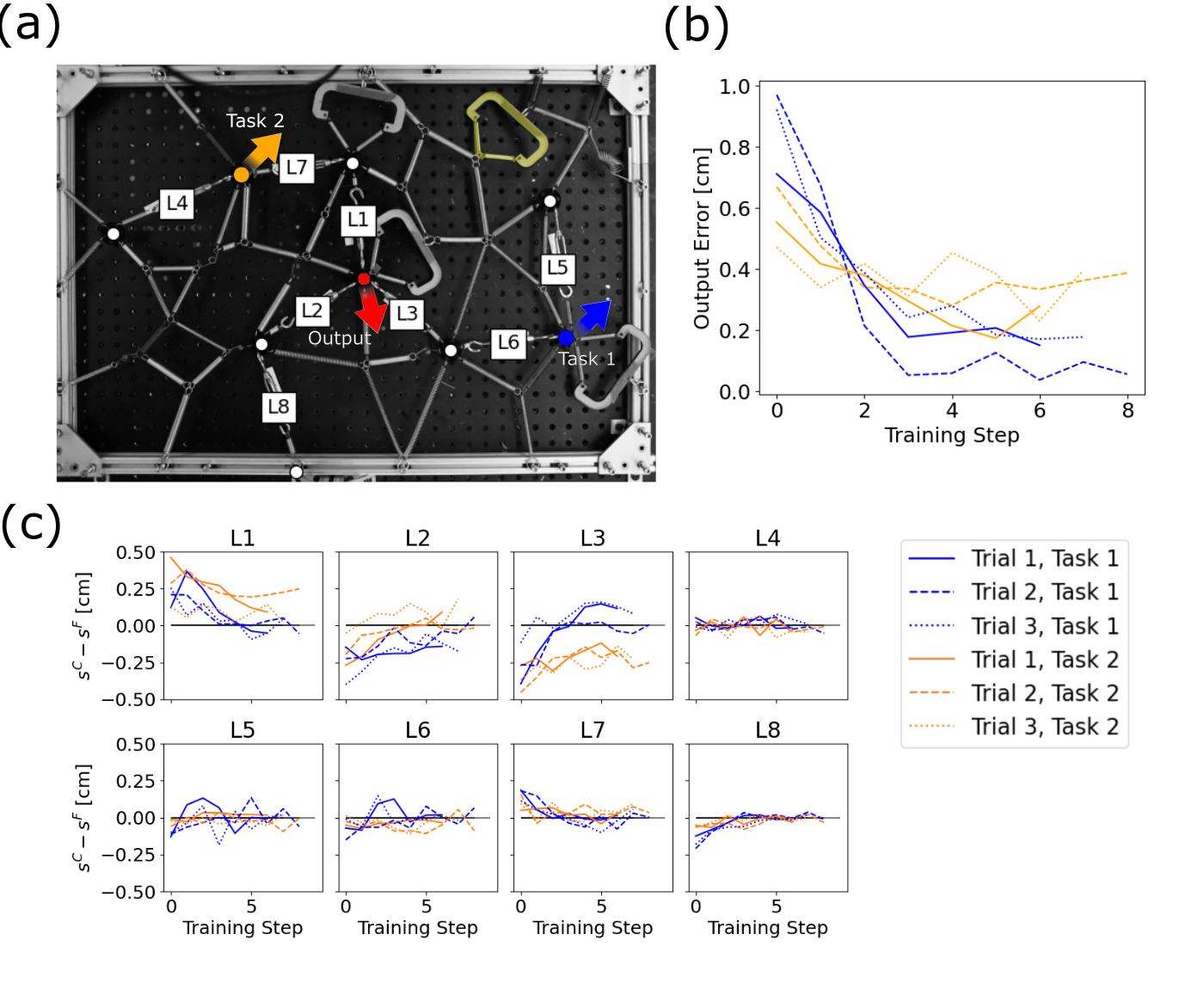}
    \caption{Training of a nonlinear elastic network for node allostery tasks.
    (a) Schematic of the network. There are \num{39} nodes and \num{59} edges, \num{8} of which have tunable rest length, labeled $L1$-$L8$. Nonlinear mechanical elements are included, such as the one highlighted in yellow which has a cubic force-displacement relationship. 
    The input nodes for task~1 and~2 are labeled in blue and orange, and the shared output node is labeled in red.
    (b) Error in position of the output node over training. The network is trained for both allostery tasks simultaneously in three separate trials. 
    In each trial, the network is initalized to a different configuration.
    Error in task~1 (blue) is reduced to a greater degree than task~2 (orange), indicating that the network can more readily learn this task.
    (c) Learning rule updates over training for each tunable edge.
    Edges with larger initial updates are more important for learning the task, and can be correlated with proximity to the output node.}
    \label{fig:nonlinear_network}
\end{figure*}

As a final example that would be highly nontrivial to model, we construct a third and much larger random network, composed of a mix of adjustable spring-turnbuckle edges and nonadjustable linear as well as nonlinear edges.
It consists of \num{59} edges and \num{39} nodes, depicted in Fig.~\ref{fig:nonlinear_network}(a). 
This network is stretched inside of a frame with \num{13} nodes fixed to the boundary.
The network is arranged in the plane perpendicular to the direction of gravity, but its internal tension is strong enough to render effects from gravity irrelevant to the experimental procedure.
Only \num{8} of these edges are tunable, labeled with ``L" in the schematic.
The majority of springs (tunable or not) in this network are linear, with a variety of native rest lengths and stiffnesses (Grainger 3HPU5).
This network also includes nonlinear springs, including torsion springs (Grainger 3HPR8, 3HPD1) and compound springs whose force-displacement relationship is cubic, such as the one highlighted in Fig.~\ref{fig:nonlinear_network}(a). 
The inclusion of nonlinear edges expands the class of potential behaviors a network can learn, and serves as an analog to a reservoir computer, in the sense that there are many non-adjustable unknown elements influencing the physical behavior of our system \cite{schrauwen2007overview}. 

We train this network simultaneously for two different node allostery tasks. 
The input nodes for each of the two tasks are labeled in blue and orange in Fig.~\ref{fig:nonlinear_network}(a), respectively.
Both tasks share the same output node, labeled in red.
Training data for each task is a single pair of two-dimensional input and output node positions, and the nodes are clamped with $\eta = 1$ using the screw holes in the optical table. 
For task~1, the input node is stretched away from the output node, and for task~2, the input is pushed towards the output.
Because our edges consist of extension springs that go slack under compression, task~1 is inherently easier to train for than task~2, as applying the input adds tension to the network, rather than releasing it. 

During training, we alternate between the two tasks at each learning step. 
Updates are made using the learning rule given in Eq.~\eqref{eq:discretelearningrule} with $\alpha = \num{0.84}$.
Training is halted when the error for each task falls below a measurement threshold, or when rest length updates average to zero over two learning steps. 
We perform training in three separate trials, each time initializing the network's tunable edge lengths to different random configurations.

The results of these trials are shown in Fig.~\ref{fig:nonlinear_network}(b).
In each case, mean output error between both tasks is reduced by a factor of about \num{5} over the course of training, which takes between \num{10} and \num{20} training steps.
The network is able to learn task~1 more quickly and accurately than task~2, finding lower-error solutions in fewer training steps. 
Among the three trials, training for task~1 achieved an average final output error of \SI{0.13}{cm}, while training for task~2 only reached \SI{0.35}{cm}. 

Fig.~\ref{fig:nonlinear_network}(c) depicts the evolution of the learning updates for each tunable edge, separated by tasks. 
Here we can see each edge's role in learning the tasks.
Edges that make larger initial updates, such as edges $L1$ and $L3$, have a stronger influence on the output node position, and would therefore change more significantly when the clamped boundary conditions are applied.
An edge's significance loosely correlates with its proximity to the output node.
We can also see how the network ``favors" learning task~1 over task~2.
While each task is trainable to zero error individually, training for both simultaneously can result in updates with respect to each task that cancel each other out, resulting in no net change in the learning degree of freedom, as is the case with $L3$ in trial~1.

\begin{figure}
    \centering
    \includegraphics[width=0.8\columnwidth]{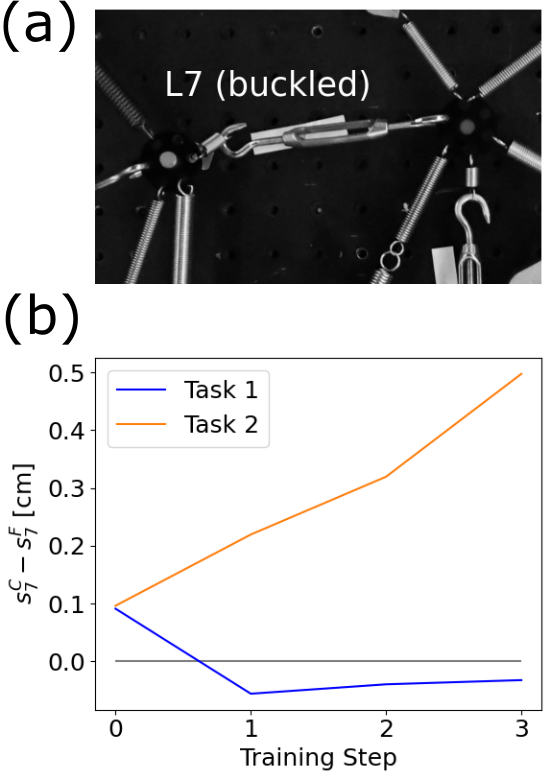}
    \caption{Instance of a failed training as a result of mechanical instability.
    (a) Photograph of tunable edge $L7$ in a buckled state as a result of the free state boundary conditions in task~2. This occurs if $L7$ is too long. 
    (b) Learning rule updates for edge $L7$ for each task. During training for task~2, the update rule steadily increases the length of $L7$, further decoupling it from the rest of the network.}
    \label{fig:failedtraining}
\end{figure}

In performing these trials, we note that training success is somewhat sensitive to the initial configuration of the network.
Fig.~\ref{fig:failedtraining} depicts a case where training of task~2 failed as a result of mechanical instability in the network. 
If edge $L7$ is initialized to be too long, then it will buckle when the free/clamped boundary conditions of task~2 are applied.
When an edge is buckled, it no longer imparts force on the network, and therefore has no effect on the position of the output node.
The learning rule will still update this edge, even though these updates are no longer correlated with the energy at that edge. 
This results in updates to $L7$ continuously increasing, rather than converging to zero.
This behavior was only observed when training task~2.

\section{Conclusion}
We have demonstrated that laboratory realizations of  mechanical networks can be trained for desired behaviors using the contrastive coupled learning algorithm. 
Through three different spring network architectures, we have shown that learning is achievable despite clear differences in the learning dynamics between simulation and experiment, or when a simulation of the physical system would be infeasible.
While functionality is limited due to their small size, these initial experiments are sufficient to demonstrate the promise of contrastive local learning in larger scale physical materials. 

The choice of a spring-turnbuckle implementation has limitations, however, and also highlights the pitfalls of our implementation. 
These edges require manual tuning, which limits the size of the network due to the labor required by the supervisor. 
Further, the requirement of a supervisor undercuts the local, decentralized nature of the learning scheme, and lessens the benefits associated with locality, such as scalability and compute time. 
Another key limitation to our approach is that each of our elastic edges is only capable of tensile forces.
As a result, under certain boundary conditions on the network, some edges can undergo mechanical instabilities such as buckling when compressed, effectively decoupling from the network, but still updating under the contrastive learning rule.
Each of our experiments were therefore performed under tension, with most edges stretched noticeably beyond their rest length. 

The choice of rest length as the learning degree of freedom also poses limitations, especially when coupled with an apparatus that can only impart force under tension. 
The presence of a nonzero rest length introduces nonlinearity in the force-displacement relation, a feature that is not present in resistive or flow networks. However, this nonlinearity is most prevalent when the spring is near its equilibrium length. 
Our extension springs learn more effectively when far above their equilibrium length, so we cannot take full advantage of this nonlinearity, or of the potential for more complex tasks enabled by nonlinearity, with this setup. See Appendix~\ref{sec:relativenodes} for more information.

Future implementations of mechanical contrastive local learning networks will include automated components that could perform local sensing and actuation to self-adjust.
Further, a mechanical element that learns by updating its stiffness rather than its rest length would allow for force as well as motion tasks.
A promising example of such an element is an electronically-controlled tunable-stiffness spring \cite{misra2023design}. 
These springs are constructed of predominantly cheap or 3D-printed parts, operate by feeding concentric coils into or out of an elastic ring to modify the stiffness, and use flex and force sensors to locally sense their elastic energies.
The modularity and ease of construction of these tunable springs should allow for significant scalability to more complex architectures and tasks.

Despite the engineering challenges, we see potential for local contrastive learning in elastic systems. The inherent nonlinearity present as a result of a nonzero rest length as well as mechanical instabilities pose opportunities to train for complex nonlinear tasks. 
Mechanical networks are capable of behaviors and material properties for which there is no analog in electrical contrastive local learning networks, such as multistability and auxeticity \cite{stern_continual_2020, reid2018auxetic}. The proof of principle we have presented here suggests that our decentralized approach could allow for learning networks to scale up.
``Smart" mechanical network materials that can learn mechanical tasks and properties locally and in real time would have far-reaching applications from soft robotics to material science.

\begin{acknowledgments}
We thank Samuel Dillavou, Ben Pisanty, Maggie Miller, Cynthia Sung, and Shivangi Misra for their helpful discussions and contributions. This work was supported by NSF grants MRSEC/DMR-1720530 (LEA, DJD) and MRSEC/DMR-2309043 (LEA, DJD), by the U.S. Department of Energy, Office of Basic Energy Sciences, Division of Materials Sciences and Engineering award DE-SC0020963 (MS), and the Simons Foundation (\# 327939 to AJL). DJD and AJL thank CCB at the Flatiron Institute, a division of the Simons Foundation, as well as the Isaac Newton Institute for Mathematical Sciences under the program ``New Statistical Physics in Living Matter" (EPSRC grant EP/R014601/1), for support and hospitality while a portion of this research was carried out. This work was initiated / performed in part at Aspen Center for Physics, which is supported by National Science Foundation grant PHY-2210452.

\end{acknowledgments}

%

\newpage
\appendix

\section{Motion Divider Learning Dynamics}
\label{sec:dynamics}
For the motion divider, we can explicitly write out and solve for the learning dynamics.  In this task, we choose an input value, $y_2$, for the position of the end node.  The system is stretched and $y_2$ is held fixed for both the free and clamped states during the entire course of training. 
The position of the middle node, $y_1$, serves as the output. It will change over the course of training, both between free and clamped states and between training steps as $L_1$ and $L_2$ evolve.
In the free state, force balance gives the position of the middle node as
\begin{equation}
    y_1^F = \frac{1}{2} y_2 + \frac{1}{2} (L_1 - L_2).
\end{equation}
Before training, this differs from the desired output value given by
\begin{equation}
    y_1^D = \frac{1}{2} y_2 + a.
\end{equation}
for whatever value the user has chosen for $a$. During training, the goal is to adjust the turnbuckle lengths so that $y_1^F \rightarrow y_1^D$, which happens when $(L_1 - L_2)/2 \rightarrow a$. The update rule to achieve this is based on comparing the free state to the clamped state, where the output node is clamped by nudge factor $0<\eta\le1$ toward the desired position
\begin{equation}
    y_1^C = \eta y_1^D + (1-\eta)y_1^F.
\end{equation}
The general discrete update rule, Eq.~(\ref{eq:discretelearningrule}), may now be evaluated by computing the node-node separations of each edge in free and clamped states using the above expressions. This gives
\begin{align}
\label{eq:discretelearningrule_eval}
    \Delta L_1 &= \alpha\eta\left(a - \frac{L_1-L_2}{2}\right) \\
    \Delta L_2 &= -\Delta L_1 \label{eq:discretelearningrule_eval_2}
\end{align}
Note that these updates are equal and opposite, and vanish when learning is achieved.  Also note that the nudge factor $\eta$ and update factor $\alpha$ equivalently affect the rate of learning. Thus, in experiment, we take $\eta=1$ and vary $\alpha$ without loss of generality. It should be emphasized that in the lab we use only Eq.~(\ref{eq:discretelearningrule}) to make the updates based on measurement of node-node separations in free and clamped states. In the lab, physics ``computes" $y_1^F$ automatically, whereas here in the appendix we additionally bring force balance to bear in order to predict learning dynamics.

To obtain differential equations from the predicted discrete update rules, one might guess that the time-derivatives of $L_1$ and $L_2$ should be proportional to the right-hand sides of 
Eqs.~\eqref{eq:discretelearningrule_eval}-\eqref{eq:discretelearningrule_eval_2}. This turns out to be true. To see, recall that the elastic energy in the clamped state must be greater than that in the free state and the difference serves as the coupled learning contrast function. For the motion divider, it can be explicitly computed using the above positions and simplifies to
\begin{align}
\label{eq:contrast_MD}
  \mathcal{C} &= U_C-U_F \\
   &= k\eta^2\left(a - \frac{L_1-L_2}{2} \right)^2,
\end{align}
Not only is this intrinsically positive, it is proportional to the loss function $\mathcal{L} = (y_1^D - y_1^F)^2$. See Appendix D for more details about when this proportionality is expected. Therefore, gradient descent on $\mathcal{C}$ is perfectly aligned with gradient descent on $\mathcal{L}$. The resulting coupled learning dynamical equations, $dL_i/dt = -\gamma d\mathcal{C}/d L_i$, simplify to
\begin{align}
\label{eq:diffeq_MD}
    \frac{\partial L_1}{\partial t} &= \nu \left( a - \frac{L_1 - L_2}{2} \right), \\
    \frac{\partial L_2}{\partial t} &= - \frac{\partial L_1}{\partial t} \label{eq:diffeq_MD_2}.
\end{align}
where $\nu = \gamma k \eta^2$ is a rate constant with units of 1/time.  These are in agreement with intuition from the discrete version above. 
The time-evolution equations, Eqs.~\eqref{eq:diffeq_MD}-\eqref{eq:diffeq_MD_2} may be directly integrated to obtain
\begin{align}
\label{eq:learning_dynamics1}
    L_1(t) &= e^{-\nu t} \left( \frac{L_1^0}{2} -  \frac{L_2^0}{2} - a \right) + a + \frac{L_1^0}{2} + \frac{L_2^0}{2}, \\
    \label{eq:learning_dynamics2}
    L_2(t) &= e^{-\nu t} \left( a - \frac{L_1^0}{2} + \frac{L_2^0}{2} \right) - a + \frac{L_1^0}{2}  + \frac{L_2^0}{2},
\end{align}
where $L_1^0$ and $L_2^0$ are the initial rest lengths of the two edges, respectively. The network stops evolving when
\begin{equation}
\label{eq:md_learned}
    \left| \frac{L_1 - L_2}{2} - a \right| \leq \epsilon,
\end{equation}
where $\epsilon$ is the desired training accuracy. Note that the analytic solutions imply that the training error
\begin{equation}
    \frac{L_1(t) - L_2(t)}{2} - a = e^{-\nu t} \left[ \frac{L_1^0 - L_2^0}{2} - a \right].
    \label{eq:md_time_evolution}
\end{equation}
decreases exponentially in time. Thus, the required duration is set by the training rate $\nu$, as well as the initial conditions and the desired accuracy.

\section{Relative and Absolute Node Positions}
\label{sec:relativenodes}

In each demonstration presented in this work, we train using inputs and outputs node positions in an external reference frame, rather than relative to their own equilibrium positions. 
This is a necessary choice for successful training in an elastic system using rest length as the learning degree of freedom.
One way to see this is to note that training in coupled learning generally only requires reference to the network under two sets of boundary conditions, the free state and the clamped state, while this training protocol would require reference to a third state, the equilibrium state. 
This is only the case because we are using rest length as the learning degree of freedom, since the equilibrium state positions change with each update step. 
When training on spring stiffnesses, the equilibrium state remains constant over training, so relative position training is allowed in this case.

We will use the simple case of a motion divider for an explicit demonstration of this constraint.
In Sections \ref{sec:motiondivider} and \ref{sec:dynamics}, we defined this system's equation of state as $y_1 = \frac{1}{2}y_2 + \frac{1}{2}(L_1 - L_2)$ and training data was defined with an input position $y_2$ and desired output position $y_1^D = \frac{1}{2} y_2 + \alpha$, with $\alpha$ as a chosen fixed value.
Both input and output positions are relative to a fixed origin, which we define as the boundary node position, depicted as the gray node in Fig.~\ref{fig:motion_divider}(a).
We could have instead defined each node's equilibrium (unstretched) positions,
\begin{align}
    y_1^0 &= L_1, \\ 
    y_2^0 &= L_1 + L_2,
\end{align}
and used the motions relative to these equilibrium positions as our training data,
\begin{align}
    y_2 &= y_2^0 + A, \\
    y_1^D &= y_1^0 + B,
\end{align}
with $A$ and $B$ as chosen fixed values.
In the free state, the target node is in position
\begin{align}
    y_1^F &= \frac{1}{2} (y_2^0 + A + L_1 - L_2) \\
     &=  L_1 + \frac{A}{2},
\end{align}
and in the clamped state, 
\begin{align}
    y_1^C &= \eta y_1^D + (1- \eta) y_1^F \\
    &= \eta (L_1 + B) + (1- \eta) (L_1 + \frac{A}{2}) \\
    &= L_1 + \frac{A}{2} + \eta (B - \frac{A}{2}).
\end{align}
Evaluating the learning rule for each edge according to Eq.~\eqref{eq:discretelearningrule}, we get
\begin{align}
    \Delta L_1 &= \alpha \eta (B - \frac{A}{2}), \\
    \Delta L_2 &= \alpha \eta (\frac{A}{2} - B).
\end{align}
The learning rule update for each edge is now a constant value, and therefore will never converge regardless of the values of $L_1$ and $L_2$. The training error, 
\begin{equation}
    y_1^D - y_1^F = B - \frac{A}{2},
\end{equation} 
is also a fixed value and does not depend on the learning degrees of freedom.

Another key difference to note is that changing an edge's rest length does significantly change the force delivered across that edge, $F = k(x-l)$, when the edge's extension is far from the equilibrium point, $x \gg l$.
Altering an edge's stiffness affects the proportionality of force to motion, but the rest length only affects the position of zero force.
Therefore, the greatest potential for change in the network's behavior by tuning rest length is to alter which edges are under extension or compression.
As we have previously stated, our turnbuckle-based apparatus is constructed such that edges buckle when $x < l$, so our experiments are typically performed such that each edge is under extension and far from its equilibrium point. 
Section~\ref{sec:nonlinear_net} and Fig.~\ref{fig:failedtraining} demonstrate how training can fail when edges buckle.
In simulation of larger networks, training for relative motions using rest length can sometimes succeed, but because the motions do not change significantly, the network cannot be detuned very far from the solution. 
Differences in node positions between the free and clamped states are therefore well below the measurement and clamping precision available to us in experiments.

\section{One-Output Symmetry Training}
\label{sec:onetarget}

As expressed in Sec.~\ref{sec:symmetrynet}, there are multiple ways to define a task in the physical degrees of freedom that is satisfied by the desired state, $L_1 = L_2$ and $L_4 = L_5$.
Such a well-defined task of this nature could be
\begin{center}
\label{eq:symmetrytask_onetarget}
    $x_{\text{target}} = 0$ for any \textit{two} $y_{\text{source}}$ values.
\end{center}
Unlike the two-output training, which only required a single data point to train, this training scheme requires that we choose two values of $y_{\text{source}}$, which we alternate between during training steps.

We train the network using this training scheme in simulation, using the equilibrium node position $y_{\text{source}}^{eq}$ and the bottom edge of the frame as the two input data points. The outputs data points are both at $x=0$, in the center of the frame. Choosing two input data points that are well separated increases the speed of learning because the nodes move a greater distance between the alternating training steps.

\begin{figure}[t!]
    \centering
    \includegraphics[width=0.45\textwidth]{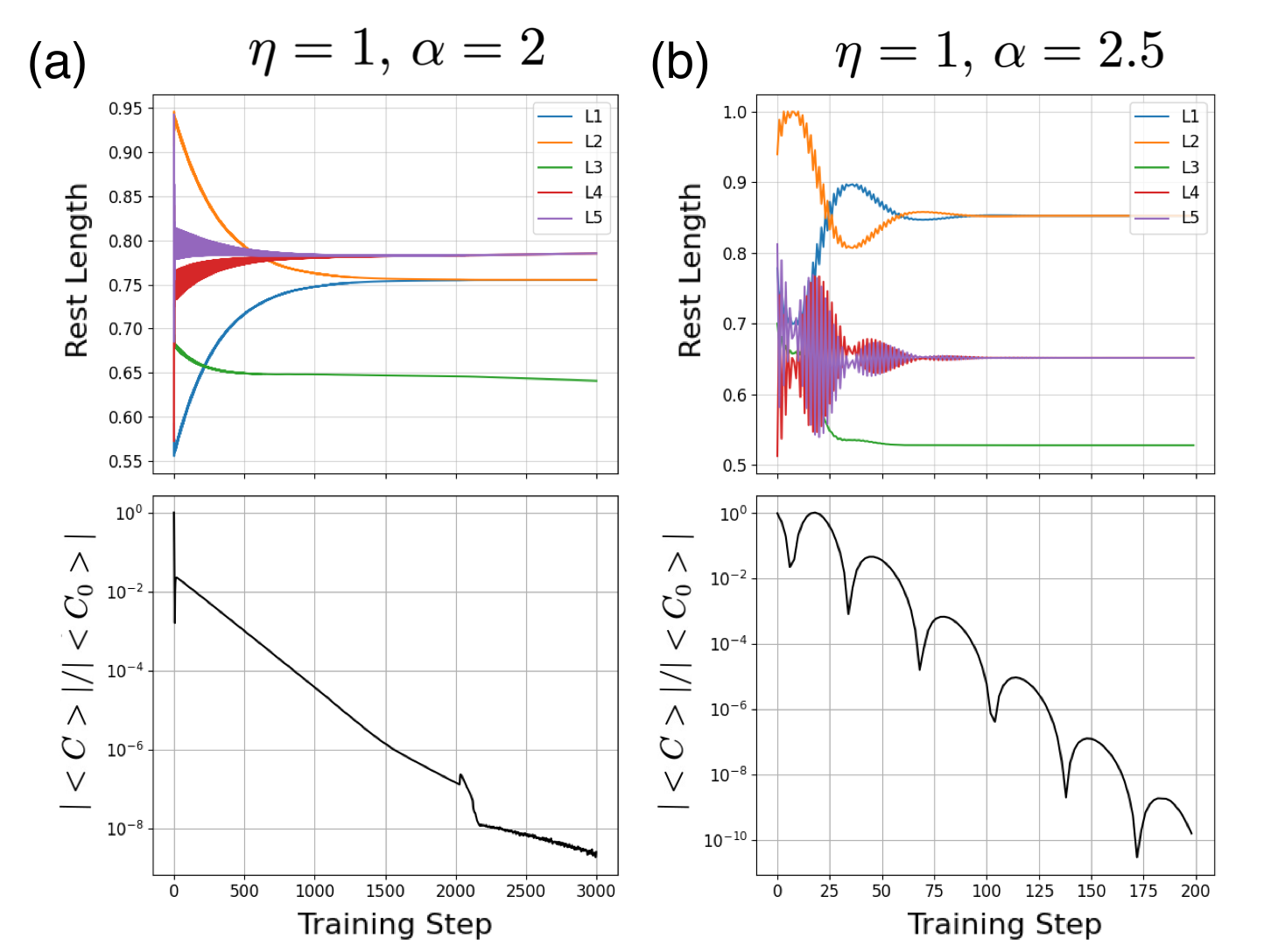}
    \caption{Simulated training of the one-output symmetry training with varying hyperparameters. Learning degrees of freedom oscillate over the course of training, and even the cost function does not decrease monotonically. Training with $\eta = \num{1}$ and $\alpha = \num{2}$ (a) takes approximately \num{2000} steps to reach the minimal cost value. Training with a high learning rate $\alpha = \num{2.5}$ (b) reduces the training time to only \num{200} but displays unstable learning dynamics.
    }
    \label{fig:oneoutput_sim}
\end{figure}

Two simulated trainings of this one-output task are shown in Fig.~\ref{fig:oneoutput_sim}. 
Both trainings use the full nudge, $\eta = 1$, and the learning rate $\alpha$ is varied.
Fig.~\ref{fig:oneoutput_sim}(a) depicts the standard training regime, where $\alpha = 2$. 
Training takes an excessively long \num{2000} steps to reach the minimum cost, with training times increasing with decreasing learning rate.
$L_4$ and $L_5$ exhibit oscillatory dynamics over the course of training, while $L_1$ and $L_2$ incrementally move towards their goal values.
Increasing the learning rate to $\alpha s = \num{2.5}$ moves us into the ``overtraining" regime, Fig.~\ref{fig:oneoutput_sim}(b), in which each learning degree of freedom can evolve \emph{past} its goal value in a single training step. 
Here, training time is reduced significantly to only \num{200} steps, but the cost function does not decrease monotonically over the course of training. 
The cost can increase by up to an order of magnitude in only a few training steps, but on average decreases by \num{10} orders of magnitude.
The rest lengths also exhibit strong oscillatory behavior, and can even switch their symmetry states between training steps.
While this type of overtraining is in principle possible to perform experimentally, the unstable learning dynamics make for a non-ideal demonstration of coupled learning's capabilities.

\section{Alignment of Coupled Learning with Gradient Descent for Elastic Networks}
\label{sec:alignment}

In this work we showed how coupled learning~\cite{stern_supervised_2021} can be used to train elastic networks for desired responses in experimental conditions. Here we discuss the coupled learning update of Eq.~\ref{eq:localsimplified} for elastic networks in more detail and discuss theoretical aspects of its experimental implementation.

Consider the node positions of the free state of the network $\mathbf{x}^F$, obtained by physical relaxation of the springs subject to inputs. By definition, the free state is a local energy minimum, so that the energy of a state in its close proximity $\mathbf{x}^F+\delta \mathbf{x}$ is

\begin{equation}
\begin{aligned}
U(\mathbf{x}^F+\delta \mathbf{x};L_i)&\approx U(\mathbf{x}^F;L_i)+ \frac{1}{2}\delta \mathbf{x}^T H(\mathbf{x}^F;L_i) \delta \mathbf{x} =\\
&= U^F + \frac{1}{2}\delta \mathbf{x}^T H^F \delta \mathbf{x}
\end{aligned}
  \label{eq:Apx1}
\end{equation}

where $H(\mathbf{x}^F;L_i)\equiv H^F = \frac{d^2 U}{\partial\mathbf{x} \partial\mathbf{x}}$ is the \emph{physical Hessian} of the network, i.e. the matrix of second derivatives of the energy with respect to node positions, evaluated at the free state.

In coupled learning, we define the clamped state with Eq.~\ref{eq:clamp}, by clamping the output nodes closer to their desired positions. The physics now minimize the network energy associated with these clamping constraints, which can be encoded using Lagrange multipliers

\begin{equation}
\begin{aligned}
U = U^F + \frac{1}{2}\delta \mathbf{x}^T H^F \delta \mathbf{x} + \lambda_O [\vec{x}^C - \vec{x}^F - \eta (\vec{x}^D - \vec{x}^F )]_O
\end{aligned}
  \label{eq:Apx2}
\end{equation}

The subscripts $O$ indicate that the Lagrange multipliers only associate with the physical degrees of freedom constrained in the output. Assuming a small nudge ($\eta \ll 1$), the clamped state minimizing Eq.~\ref{eq:Apx2} is very close to the free state and belongs to the same energy basin. In this regime, one can solve for the clamped state given the free state positions $\mathbf{x}^F$ and the clamping constraints. 

To do that, we can recast Eq.~\ref{eq:Apx2} by defining an extended `position' vector $\mathbf{y}\equiv[\mathbf{x}, \lambda_O]^T$, and a clamping `force' vector $\mathbf{f}^O\equiv  [\mathbf{0}, x^D_O - x^F_O]$. Note that when no clamping force is applied $\lambda_O=0$ and the free state is given by $\mathbf{y}^F=[\mathbf{x}^F, 0_O]^T$. We further construct an output matrix $S_O$ with $N$ (physical degrees of freedom) columns and a number of rows equal to the number of output degrees of freedom $N_O$. Each row in this matrix has a single non-vanishing component of value $1$, at the index associated with the corresponding output physical degree of freedom. With this, we can define a \emph{clamped physical Hessian} $H^C$

\begin{equation}
\begin{aligned}
H^C=\begin{pmatrix}
H^F & S_O^T  \\
S_O & 0 \\
\end{pmatrix}.
\end{aligned}
\label{eq:Apx3}
\end{equation}

Using these definitions, we can write the physical optimization problem of Eq.~\ref{eq:Apx2} and its solution (see~\cite{stern2024physical}) as

\begin{equation}
\begin{aligned}
U &= U^F + \frac{1}{2}\delta \mathbf{y}^T H^C \delta \mathbf{y} - \eta \delta \mathbf{y}^T \mathbf{f}^O \\
\mathbf{y}^C &= \mathbf{y}^F + \eta (H^C)^{-1} \mathbf{f}^O \\ 
U^C &= U^F - \frac{1}{2}\eta^2 (x^D-x^F)_O^T(H^C)^{-1}_{OO} (x^D-x^F)_O
\end{aligned}
  \label{eq:Apx4}
\end{equation}

with $(H^C)^{-1}$ the inverse of the clamped physical Hessian, and $(H^C)^{-1}_{OO}$ is the negative definite bottom right block of this inverse matrix.

Using the contrastive learning framework, the coupled learning rule is given by the difference between the clamped and free energies, differentiated with respect to the learning degrees of freedom (rest lengths):

\begin{equation}
\begin{aligned}
\Delta L_i (CL) &\sim \eta^{-1}\frac{\partial(U^F - U^C)}{\partial_{L_i}} \approx \\
&\approx (x^D-x^F)^T_O (H^C)^{-1}_{OO} \frac{d}{dL_i} x^F_O
\end{aligned}
  \label{eq:Apx5}
\end{equation}

A more standard path to training the elastic networks is to choose the learning rule according to gradient descent, i.e., by computing the gradient of the learning cost function $C=\frac{1}{2}(x^D-x^F)_O^2$ with respect to the learning degrees of freedom. It is easy to verify that the gradient descent modification is given by

\begin{equation}
\begin{aligned}
\Delta L_i (GD) &\sim -\frac{dC}{dL_i} = \\
&= - (x^D-x^F)^T_O \frac{d}{dL_i} x^F_O
\end{aligned}
  \label{eq:Apx6}
\end{equation}

One can see the similarity between the local learning rule afforded by coupled learning and the global gradient computation. The difference between them stems from the insertion of the negative definite matrix $(H^C)^{-1}_{OO}$. In the special case where there is just a single output degree of freedom, this matrix is just a negative scalar, and the coupled learning rule is proportional to that obtained by gradient descent, as discussed for such a case in Appendix A. However, this is no longer the case more complicated tasks with multiple output degrees of freedom. In such cases the two rules of Eqs.~\ref{eq:Apx5}-\ref{eq:Apx6} are no longer proportional, as the matrix $(H^C)^{-1}_{OO}$ is in general not proportional to identity matrix. The modifications vector would then be misaligned, and in extreme cases one might even have a negative projection on the other (in which cases coupled learning would actually locally \emph{increase} the error). However, we found numerically that in high likelihood, the two learning rules are correlated and point in similar directions in the space of learning degrees of freedom~\cite{stern_supervised_2021}. Even if the alignment between them is locally poor, a few steps of the learning process in the wrong direction produce a brief ``unlearning spike", during which the error grows, but then the two rules align again and effective learning resumes. We have never numerically observed coupled learning fail by being consistently misaligned with the cost gradient, though we cannot rule out the existence of such cases.

Another important qualification to the above considerations is that they are only strictly valid in the linear regime around the free state configuration. The contrastive learning (and coupled learning theory) assumes that the clamped state is infinitesimally close to the free state, i.e. that small nudges ($\eta\ll 1$) are applied in the clamping. In that limit one can treat the difference between the free and clamped energies as a derivative, and a connection can be drawn between the contrast and the cost function one wishes to minimize. In this $\eta\ll 1$ limit, the free and clamped states are obviously very similar, as the difference between them scale with $\eta$ (Eq.~\ref{eq:Apx4}). In systems with thermal noise or uncertain measurements, one cannot use arbitrarily small nudges, as any noise in the measurement would be much larger than the contrast between the two states. In our experiments we thus use large nudges $\eta\approx 1$, which introduces various theoretical complications. For one, the theoretical relationships drawn between the local learning rule and the cost gradient are only valid to first order in $\eta$ so that large nudges worsen the approximation of the cost gradient afforded by our local learning rule. Even worse, as elastic networks are non-linear, clamping with large nudges can move the configuration into a different basin in the energy landscape (compared to the free state). When the clamped state can no longer be treated as a perturbation around the free state, the formal justification for contrastive learning is lost, and we can no longer guarantee learning convergence. Such cases often also lead to the aforementioned ``unlearning spikes" and growing error. In practice, though, we find that learning in our system is successful in numerical and experimental settings even in these conditions; the network eventually discovers a configuration of the learning degrees of freedom for which the free and clamped states occur in the same energy basin, and coupled learning is successful from there.


\begin{thebibliography}{40}%
\makeatletter
\providecommand \@ifxundefined [1]{%
 \@ifx{#1\undefined}
}%
\providecommand \@ifnum [1]{%
 \ifnum #1\expandafter \@firstoftwo
 \else \expandafter \@secondoftwo
 \fi
}%
\providecommand \@ifx [1]{%
 \ifx #1\expandafter \@firstoftwo
 \else \expandafter \@secondoftwo
 \fi
}%
\providecommand \natexlab [1]{#1}%
\providecommand \enquote  [1]{``#1''}%
\providecommand \bibnamefont  [1]{#1}%
\providecommand \bibfnamefont [1]{#1}%
\providecommand \citenamefont [1]{#1}%
\providecommand \href@noop [0]{\@secondoftwo}%
\providecommand \href [0]{\begingroup \@sanitize@url \@href}%
\providecommand \@href[1]{\@@startlink{#1}\@@href}%
\providecommand \@@href[1]{\endgroup#1\@@endlink}%
\providecommand \@sanitize@url [0]{\catcode `\\12\catcode `\$12\catcode
  `\&12\catcode `\#12\catcode `\^12\catcode `\_12\catcode `\%12\relax}%
\providecommand \@@startlink[1]{}%
\providecommand \@@endlink[0]{}%
\providecommand \url  [0]{\begingroup\@sanitize@url \@url }%
\providecommand \@url [1]{\endgroup\@href {#1}{\urlprefix }}%
\providecommand \urlprefix  [0]{URL }%
\providecommand \Eprint [0]{\href }%
\providecommand \doibase [0]{https://doi.org/}%
\providecommand \selectlanguage [0]{\@gobble}%
\providecommand \bibinfo  [0]{\@secondoftwo}%
\providecommand \bibfield  [0]{\@secondoftwo}%
\providecommand \translation [1]{[#1]}%
\providecommand \BibitemOpen [0]{}%
\providecommand \bibitemStop [0]{}%
\providecommand \bibitemNoStop [0]{.\EOS\space}%
\providecommand \EOS [0]{\spacefactor3000\relax}%
\providecommand \BibitemShut  [1]{\csname bibitem#1\endcsname}%
\let\auto@bib@innerbib\@empty
\bibitem [{\citenamefont {LeCun}\ \emph {et~al.}(2015)\citenamefont {LeCun},
  \citenamefont {Bengio},\ and\ \citenamefont {Hinton}}]{lecun_deep_2015}%
  \BibitemOpen
  \bibfield  {author} {\bibinfo {author} {\bibfnamefont {Y.}~\bibnamefont
  {LeCun}}, \bibinfo {author} {\bibfnamefont {Y.}~\bibnamefont {Bengio}},\ and\
  \bibinfo {author} {\bibfnamefont {G.}~\bibnamefont {Hinton}},\ }\bibfield
  {title} {\bibinfo {title} {Deep learning},\ }\href
  {https://doi.org/10.1038/nature14539} {\bibfield  {journal} {\bibinfo
  {journal} {Nature}\ }\textbf {\bibinfo {volume} {521}},\ \bibinfo {pages}
  {436} (\bibinfo {year} {2015})}\BibitemShut {NoStop}%
\bibitem [{\citenamefont {Rojas}(1996)}]{Rojas1996}%
  \BibitemOpen
  \bibfield  {author} {\bibinfo {author} {\bibfnamefont {R.}~\bibnamefont
  {Rojas}},\ }\bibinfo {title} {The backpropagation algorithm},\ in\ \href
  {https://doi.org/10.1007/978-3-642-61068-4_7} {\emph {\bibinfo {booktitle}
  {Neural Networks: A Systematic Introduction}}}\ (\bibinfo  {publisher}
  {Springer Berlin Heidelberg},\ \bibinfo {address} {Berlin, Heidelberg},\
  \bibinfo {year} {1996})\ pp.\ \bibinfo {pages} {149--182}\BibitemShut
  {NoStop}%
\bibitem [{\citenamefont {Rocks}\ \emph {et~al.}(2017)\citenamefont {Rocks},
  \citenamefont {Pashine}, \citenamefont {Bischofberger}, \citenamefont
  {Goodrich}, \citenamefont {Liu},\ and\ \citenamefont
  {Nagel}}]{rocks_designing_2017}%
  \BibitemOpen
  \bibfield  {author} {\bibinfo {author} {\bibfnamefont {J.~W.}\ \bibnamefont
  {Rocks}}, \bibinfo {author} {\bibfnamefont {N.}~\bibnamefont {Pashine}},
  \bibinfo {author} {\bibfnamefont {I.}~\bibnamefont {Bischofberger}}, \bibinfo
  {author} {\bibfnamefont {C.~P.}\ \bibnamefont {Goodrich}}, \bibinfo {author}
  {\bibfnamefont {A.~J.}\ \bibnamefont {Liu}},\ and\ \bibinfo {author}
  {\bibfnamefont {S.~R.}\ \bibnamefont {Nagel}},\ }\bibfield  {title} {\bibinfo
  {title} {Designing allostery-inspired response in mechanical networks},\
  }\href {https://doi.org/10.1073/pnas.1612139114} {\bibfield  {journal}
  {\bibinfo  {journal} {Proceedings of the National Academy of Sciences}\
  }\textbf {\bibinfo {volume} {114}},\ \bibinfo {pages} {2520} (\bibinfo {year}
  {2017})}\BibitemShut {NoStop}%
\bibitem [{\citenamefont {Reid}\ \emph {et~al.}(2018)\citenamefont {Reid},
  \citenamefont {Pashine}, \citenamefont {Wozniak}, \citenamefont {Jaeger},
  \citenamefont {Liu}, \citenamefont {Nagel},\ and\ \citenamefont
  {de~Pablo}}]{reid2018auxetic}%
  \BibitemOpen
  \bibfield  {author} {\bibinfo {author} {\bibfnamefont {D.~R.}\ \bibnamefont
  {Reid}}, \bibinfo {author} {\bibfnamefont {N.}~\bibnamefont {Pashine}},
  \bibinfo {author} {\bibfnamefont {J.~M.}\ \bibnamefont {Wozniak}}, \bibinfo
  {author} {\bibfnamefont {H.~M.}\ \bibnamefont {Jaeger}}, \bibinfo {author}
  {\bibfnamefont {A.~J.}\ \bibnamefont {Liu}}, \bibinfo {author} {\bibfnamefont
  {S.~R.}\ \bibnamefont {Nagel}},\ and\ \bibinfo {author} {\bibfnamefont
  {J.~J.}\ \bibnamefont {de~Pablo}},\ }\bibfield  {title} {\bibinfo {title}
  {Auxetic metamaterials from disordered networks},\ }\href@noop {} {\bibfield
  {journal} {\bibinfo  {journal} {Proceedings of the National Academy of
  Sciences}\ }\textbf {\bibinfo {volume} {115}},\ \bibinfo {pages} {E1384}
  (\bibinfo {year} {2018})}\BibitemShut {NoStop}%
\bibitem [{\citenamefont {Goodrich}\ \emph {et~al.}(2015)\citenamefont
  {Goodrich}, \citenamefont {Liu},\ and\ \citenamefont
  {Nagel}}]{goodrich_principle_2015}%
  \BibitemOpen
  \bibfield  {author} {\bibinfo {author} {\bibfnamefont {C.~P.}\ \bibnamefont
  {Goodrich}}, \bibinfo {author} {\bibfnamefont {A.~J.}\ \bibnamefont {Liu}},\
  and\ \bibinfo {author} {\bibfnamefont {S.~R.}\ \bibnamefont {Nagel}},\
  }\bibfield  {title} {\bibinfo {title} {The {{Principle}} of {{Independent
  Bond-Level Response}}: {{Tuning}} by {{Pruning}} to {{Exploit Disorder}} for
  {{Global Behavior}}},\ }\href
  {https://doi.org/10.1103/PhysRevLett.114.225501} {\bibfield  {journal}
  {\bibinfo  {journal} {Physical Review Letters}\ }\textbf {\bibinfo {volume}
  {114}},\ \bibinfo {pages} {225501} (\bibinfo {year} {2015})}\BibitemShut
  {NoStop}%
\bibitem [{\citenamefont {Wright}\ \emph {et~al.}(2022)\citenamefont {Wright},
  \citenamefont {Onodera}, \citenamefont {Stein}, \citenamefont {Wang},
  \citenamefont {Schachter}, \citenamefont {Hu},\ and\ \citenamefont
  {McMahon}}]{wright_deep_2022}%
  \BibitemOpen
  \bibfield  {author} {\bibinfo {author} {\bibfnamefont {L.~G.}\ \bibnamefont
  {Wright}}, \bibinfo {author} {\bibfnamefont {T.}~\bibnamefont {Onodera}},
  \bibinfo {author} {\bibfnamefont {M.~M.}\ \bibnamefont {Stein}}, \bibinfo
  {author} {\bibfnamefont {T.}~\bibnamefont {Wang}}, \bibinfo {author}
  {\bibfnamefont {D.~T.}\ \bibnamefont {Schachter}}, \bibinfo {author}
  {\bibfnamefont {Z.}~\bibnamefont {Hu}},\ and\ \bibinfo {author}
  {\bibfnamefont {P.~L.}\ \bibnamefont {McMahon}},\ }\bibfield  {title}
  {\bibinfo {title} {Deep physical neural networks trained with
  backpropagation},\ }\href {https://doi.org/10.1038/s41586-021-04223-6}
  {\bibfield  {journal} {\bibinfo  {journal} {Nature}\ }\textbf {\bibinfo
  {volume} {601}},\ \bibinfo {pages} {549} (\bibinfo {year}
  {2022})}\BibitemShut {NoStop}%
\bibitem [{\citenamefont {Lee}\ \emph {et~al.}(2022)\citenamefont {Lee},
  \citenamefont {Mulder},\ and\ \citenamefont {Hopkins}}]{lee2022mechanical}%
  \BibitemOpen
  \bibfield  {author} {\bibinfo {author} {\bibfnamefont {R.~H.}\ \bibnamefont
  {Lee}}, \bibinfo {author} {\bibfnamefont {E.~A.}\ \bibnamefont {Mulder}},\
  and\ \bibinfo {author} {\bibfnamefont {J.~B.}\ \bibnamefont {Hopkins}},\
  }\bibfield  {title} {\bibinfo {title} {Mechanical neural networks:
  Architected materials that learn behaviors},\ }\href@noop {} {\bibfield
  {journal} {\bibinfo  {journal} {Science Robotics}\ }\textbf {\bibinfo
  {volume} {7}},\ \bibinfo {pages} {eabq7278} (\bibinfo {year}
  {2022})}\BibitemShut {NoStop}%
\bibitem [{\citenamefont {Stern}\ and\ \citenamefont
  {Murugan}(2023)}]{stern2023learning}%
  \BibitemOpen
  \bibfield  {author} {\bibinfo {author} {\bibfnamefont {M.}~\bibnamefont
  {Stern}}\ and\ \bibinfo {author} {\bibfnamefont {A.}~\bibnamefont
  {Murugan}},\ }\bibfield  {title} {\bibinfo {title} {Learning without neurons
  in physical systems},\ }\href
  {https://doi.org/10.1146/annurev-conmatphys-040821-113439} {\bibfield
  {journal} {\bibinfo  {journal} {Annual Review of Condensed Matter Physics}\
  }\textbf {\bibinfo {volume} {14}},\ \bibinfo {pages} {417} (\bibinfo {year}
  {2023})}\BibitemShut {NoStop}%
\bibitem [{\citenamefont {Den{\`e}ve}\ \emph {et~al.}(2017)\citenamefont
  {Den{\`e}ve}, \citenamefont {Alemi},\ and\ \citenamefont
  {Bourdoukan}}]{deneve_brain_2017}%
  \BibitemOpen
  \bibfield  {author} {\bibinfo {author} {\bibfnamefont {S.}~\bibnamefont
  {Den{\`e}ve}}, \bibinfo {author} {\bibfnamefont {A.}~\bibnamefont {Alemi}},\
  and\ \bibinfo {author} {\bibfnamefont {R.}~\bibnamefont {Bourdoukan}},\
  }\bibfield  {title} {\bibinfo {title} {The {{Brain}} as an {{Efficient}} and
  {{Robust Adaptive Learner}}},\ }\href
  {https://doi.org/10.1016/j.neuron.2017.05.016} {\bibfield  {journal}
  {\bibinfo  {journal} {Neuron}\ }\textbf {\bibinfo {volume} {94}},\ \bibinfo
  {pages} {969} (\bibinfo {year} {2017})}\BibitemShut {NoStop}%
\bibitem [{\citenamefont {Sengupta}\ and\ \citenamefont
  {Stemmler}(2014)}]{sengupta_power_2014}%
  \BibitemOpen
  \bibfield  {author} {\bibinfo {author} {\bibfnamefont {B.}~\bibnamefont
  {Sengupta}}\ and\ \bibinfo {author} {\bibfnamefont {M.~B.}\ \bibnamefont
  {Stemmler}},\ }\bibfield  {title} {\bibinfo {title} {Power {{Consumption
  During Neuronal Computation}}},\ }\href
  {https://doi.org/10.1109/JPROC.2014.2307755} {\bibfield  {journal} {\bibinfo
  {journal} {Proceedings of the IEEE}\ }\textbf {\bibinfo {volume} {102}},\
  \bibinfo {pages} {738} (\bibinfo {year} {2014})}\BibitemShut {NoStop}%
\bibitem [{\citenamefont {Stern}\ \emph
  {et~al.}(2024{\natexlab{a}})\citenamefont {Stern}, \citenamefont {Dillavou},
  \citenamefont {Jayaraman}, \citenamefont {Durian},\ and\ \citenamefont
  {Liu}}]{stern2024training}%
  \BibitemOpen
  \bibfield  {author} {\bibinfo {author} {\bibfnamefont {M.}~\bibnamefont
  {Stern}}, \bibinfo {author} {\bibfnamefont {S.}~\bibnamefont {Dillavou}},
  \bibinfo {author} {\bibfnamefont {D.}~\bibnamefont {Jayaraman}}, \bibinfo
  {author} {\bibfnamefont {D.~J.}\ \bibnamefont {Durian}},\ and\ \bibinfo
  {author} {\bibfnamefont {A.~J.}\ \bibnamefont {Liu}},\ }\bibfield  {title}
  {\bibinfo {title} {Training self-learning circuits for power-efficient
  solutions},\ }\href@noop {} {\bibfield  {journal} {\bibinfo  {journal} {APL
  Machine Learning}\ }\textbf {\bibinfo {volume} {2}} (\bibinfo {year}
  {2024}{\natexlab{a}})}\BibitemShut {NoStop}%
\bibitem [{\citenamefont {John}\ \emph {et~al.}(2020)\citenamefont {John},
  \citenamefont {Tiwari}, \citenamefont {Patdillah}, \citenamefont {Kulkarni},
  \citenamefont {Tiwari}, \citenamefont {Basu}, \citenamefont {Bose},
  \citenamefont {Ankit}, \citenamefont {Yu}, \citenamefont {Nirmal} \emph
  {et~al.}}]{john2020self}%
  \BibitemOpen
  \bibfield  {author} {\bibinfo {author} {\bibfnamefont {R.~A.}\ \bibnamefont
  {John}}, \bibinfo {author} {\bibfnamefont {N.}~\bibnamefont {Tiwari}},
  \bibinfo {author} {\bibfnamefont {M.~I.~B.}\ \bibnamefont {Patdillah}},
  \bibinfo {author} {\bibfnamefont {M.~R.}\ \bibnamefont {Kulkarni}}, \bibinfo
  {author} {\bibfnamefont {N.}~\bibnamefont {Tiwari}}, \bibinfo {author}
  {\bibfnamefont {J.}~\bibnamefont {Basu}}, \bibinfo {author} {\bibfnamefont
  {S.~K.}\ \bibnamefont {Bose}}, \bibinfo {author} {\bibnamefont {Ankit}},
  \bibinfo {author} {\bibfnamefont {C.~J.}\ \bibnamefont {Yu}}, \bibinfo
  {author} {\bibfnamefont {A.}~\bibnamefont {Nirmal}}, \emph {et~al.},\
  }\bibfield  {title} {\bibinfo {title} {Self healable neuromorphic
  memtransistor elements for decentralized sensory signal processing in
  robotics},\ }\href@noop {} {\bibfield  {journal} {\bibinfo  {journal} {Nature
  communications}\ }\textbf {\bibinfo {volume} {11}},\ \bibinfo {pages} {4030}
  (\bibinfo {year} {2020})}\BibitemShut {NoStop}%
\bibitem [{\citenamefont {McGovern}\ \emph {et~al.}(2019)\citenamefont
  {McGovern}, \citenamefont {Moosa}, \citenamefont {Jehi}, \citenamefont
  {Busch}, \citenamefont {Ferguson}, \citenamefont {Gupta}, \citenamefont
  {Gonzalez-Martinez}, \citenamefont {Wyllie}, \citenamefont {Najm},\ and\
  \citenamefont {Bingaman}}]{mcgovern_hemispherectomy_2019}%
  \BibitemOpen
  \bibfield  {author} {\bibinfo {author} {\bibfnamefont {R.~A.}\ \bibnamefont
  {McGovern}}, \bibinfo {author} {\bibfnamefont {A.~N.~V.}\ \bibnamefont
  {Moosa}}, \bibinfo {author} {\bibfnamefont {L.}~\bibnamefont {Jehi}},
  \bibinfo {author} {\bibfnamefont {R.}~\bibnamefont {Busch}}, \bibinfo
  {author} {\bibfnamefont {L.}~\bibnamefont {Ferguson}}, \bibinfo {author}
  {\bibfnamefont {A.}~\bibnamefont {Gupta}}, \bibinfo {author} {\bibfnamefont
  {J.}~\bibnamefont {Gonzalez-Martinez}}, \bibinfo {author} {\bibfnamefont
  {E.}~\bibnamefont {Wyllie}}, \bibinfo {author} {\bibfnamefont
  {I.}~\bibnamefont {Najm}},\ and\ \bibinfo {author} {\bibfnamefont {W.~E.}\
  \bibnamefont {Bingaman}},\ }\bibfield  {title} {\bibinfo {title}
  {Hemispherectomy in adults and adolescents: {{Seizure}} and functional
  outcomes in 47 patients},\ }\href {https://doi.org/10.1111/epi.16378}
  {\bibfield  {journal} {\bibinfo  {journal} {Epilepsia}\ }\textbf {\bibinfo
  {volume} {60}},\ \bibinfo {pages} {2416} (\bibinfo {year}
  {2019})}\BibitemShut {NoStop}%
\bibitem [{\citenamefont {Dillavou}\ \emph {et~al.}(2022)\citenamefont
  {Dillavou}, \citenamefont {Stern}, \citenamefont {Liu},\ and\ \citenamefont
  {Durian}}]{dillavou2022demonstration}%
  \BibitemOpen
  \bibfield  {author} {\bibinfo {author} {\bibfnamefont {S.}~\bibnamefont
  {Dillavou}}, \bibinfo {author} {\bibfnamefont {M.}~\bibnamefont {Stern}},
  \bibinfo {author} {\bibfnamefont {A.~J.}\ \bibnamefont {Liu}},\ and\ \bibinfo
  {author} {\bibfnamefont {D.~J.}\ \bibnamefont {Durian}},\ }\bibfield  {title}
  {\bibinfo {title} {Demonstration of decentralized physics-driven learning},\
  }\href@noop {} {\bibfield  {journal} {\bibinfo  {journal} {Physical Review
  Applied}\ }\textbf {\bibinfo {volume} {18}},\ \bibinfo {pages} {014040}
  (\bibinfo {year} {2022})}\BibitemShut {NoStop}%
\bibitem [{\citenamefont {de~Vries}(2023)}]{de2023growing}%
  \BibitemOpen
  \bibfield  {author} {\bibinfo {author} {\bibfnamefont {A.}~\bibnamefont
  {de~Vries}},\ }\bibfield  {title} {\bibinfo {title} {The growing energy
  footprint of artificial intelligence},\ }\href@noop {} {\bibfield  {journal}
  {\bibinfo  {journal} {Joule}\ }\textbf {\bibinfo {volume} {7}},\ \bibinfo
  {pages} {2191} (\bibinfo {year} {2023})}\BibitemShut {NoStop}%
\bibitem [{\citenamefont {Pashine}\ \emph {et~al.}(2019)\citenamefont
  {Pashine}, \citenamefont {Hexner}, \citenamefont {Liu},\ and\ \citenamefont
  {Nagel}}]{pashine_directed_2019}%
  \BibitemOpen
  \bibfield  {author} {\bibinfo {author} {\bibfnamefont {N.}~\bibnamefont
  {Pashine}}, \bibinfo {author} {\bibfnamefont {D.}~\bibnamefont {Hexner}},
  \bibinfo {author} {\bibfnamefont {A.~J.}\ \bibnamefont {Liu}},\ and\ \bibinfo
  {author} {\bibfnamefont {S.~R.}\ \bibnamefont {Nagel}},\ }\bibfield  {title}
  {\bibinfo {title} {Directed aging, memory, and nature's greed},\ }\href
  {https://doi.org/10.1126/sciadv.aax4215} {\bibfield  {journal} {\bibinfo
  {journal} {Science Advances}\ }\textbf {\bibinfo {volume} {5}},\ \bibinfo
  {pages} {eaax4215} (\bibinfo {year} {2019})}\BibitemShut {NoStop}%
\bibitem [{\citenamefont {Hexner}\ \emph
  {et~al.}(2020{\natexlab{a}})\citenamefont {Hexner}, \citenamefont {Pashine},
  \citenamefont {Liu},\ and\ \citenamefont {Nagel}}]{hexner_effect_2020}%
  \BibitemOpen
  \bibfield  {author} {\bibinfo {author} {\bibfnamefont {D.}~\bibnamefont
  {Hexner}}, \bibinfo {author} {\bibfnamefont {N.}~\bibnamefont {Pashine}},
  \bibinfo {author} {\bibfnamefont {A.~J.}\ \bibnamefont {Liu}},\ and\ \bibinfo
  {author} {\bibfnamefont {S.~R.}\ \bibnamefont {Nagel}},\ }\bibfield  {title}
  {\bibinfo {title} {Effect of directed aging on nonlinear elasticity and
  memory formation in a material},\ }\href
  {https://doi.org/10.1103/PhysRevResearch.2.043231} {\bibfield  {journal}
  {\bibinfo  {journal} {Physical Review Research}\ }\textbf {\bibinfo {volume}
  {2}},\ \bibinfo {pages} {043231} (\bibinfo {year}
  {2020}{\natexlab{a}})}\BibitemShut {NoStop}%
\bibitem [{\citenamefont {Hexner}\ \emph
  {et~al.}(2020{\natexlab{b}})\citenamefont {Hexner}, \citenamefont {Liu},\
  and\ \citenamefont {Nagel}}]{hexner_periodic_2020}%
  \BibitemOpen
  \bibfield  {author} {\bibinfo {author} {\bibfnamefont {D.}~\bibnamefont
  {Hexner}}, \bibinfo {author} {\bibfnamefont {A.~J.}\ \bibnamefont {Liu}},\
  and\ \bibinfo {author} {\bibfnamefont {S.~R.}\ \bibnamefont {Nagel}},\
  }\bibfield  {title} {\bibinfo {title} {Periodic training of creeping
  solids},\ }\href {https://doi.org/10.1073/pnas.1922847117} {\bibfield
  {journal} {\bibinfo  {journal} {Proceedings of the National Academy of
  Sciences}\ }\textbf {\bibinfo {volume} {117}},\ \bibinfo {pages} {31690}
  (\bibinfo {year} {2020}{\natexlab{b}})}\BibitemShut {NoStop}%
\bibitem [{\citenamefont {Movellan}(1991)}]{movellan_contrastive_1991}%
  \BibitemOpen
  \bibfield  {author} {\bibinfo {author} {\bibfnamefont {J.~R.}\ \bibnamefont
  {Movellan}},\ }\bibfield  {title} {\bibinfo {title} {Contrastive {{Hebbian
  Learning}} in the {{Continuous Hopfield Model}}},\ }in\ \href
  {https://doi.org/10.1016/B978-1-4832-1448-1.50007-X} {\emph {\bibinfo
  {booktitle} {Connectionist {{Models}}}}},\ \bibinfo {editor} {edited by\
  \bibinfo {editor} {\bibfnamefont {D.~S.}\ \bibnamefont {Touretzky}}, \bibinfo
  {editor} {\bibfnamefont {J.~L.}\ \bibnamefont {Elman}}, \bibinfo {editor}
  {\bibfnamefont {T.~J.}\ \bibnamefont {Sejnowski}},\ and\ \bibinfo {editor}
  {\bibfnamefont {G.~E.}\ \bibnamefont {Hinton}}}\ (\bibinfo  {publisher}
  {{Morgan Kaufmann}},\ \bibinfo {year} {1991})\ pp.\ \bibinfo {pages}
  {10--17}\BibitemShut {NoStop}%
\bibitem [{\citenamefont {Pashine}(2021)}]{pashine_local_2021}%
  \BibitemOpen
  \bibfield  {author} {\bibinfo {author} {\bibfnamefont {N.}~\bibnamefont
  {Pashine}},\ }\bibfield  {title} {\bibinfo {title} {Local rules for
  fabricating allosteric networks},\ }\href
  {https://doi.org/10.1103/PhysRevMaterials.5.065607} {\bibfield  {journal}
  {\bibinfo  {journal} {Physical Review Materials}\ }\textbf {\bibinfo {volume}
  {5}},\ \bibinfo {pages} {065607} (\bibinfo {year} {2021})}\BibitemShut
  {NoStop}%
\bibitem [{\citenamefont {Scellier}\ and\ \citenamefont
  {Bengio}(2017)}]{scellier_equilibrium_2017}%
  \BibitemOpen
  \bibfield  {author} {\bibinfo {author} {\bibfnamefont {B.}~\bibnamefont
  {Scellier}}\ and\ \bibinfo {author} {\bibfnamefont {Y.}~\bibnamefont
  {Bengio}},\ }\bibfield  {title} {\bibinfo {title} {Equilibrium
  {{Propagation}}: {{Bridging}} the {{Gap}} between {{Energy-Based Models}} and
  {{Backpropagation}}},\ }\bibfield  {journal} {\bibinfo  {journal} {Frontiers
  in Computational Neuroscience}\ }\textbf {\bibinfo {volume} {11}},\ \href
  {https://doi.org/10.3389/fncom.2017.00024} {10.3389/fncom.2017.00024}
  (\bibinfo {year} {2017})\BibitemShut {NoStop}%
\bibitem [{\citenamefont {Kendall}\ \emph {et~al.}(2020)\citenamefont
  {Kendall}, \citenamefont {Pantone}, \citenamefont {Manickavasagam},
  \citenamefont {Bengio},\ and\ \citenamefont
  {Scellier}}]{kendall_training_2020}%
  \BibitemOpen
  \bibfield  {author} {\bibinfo {author} {\bibfnamefont {J.}~\bibnamefont
  {Kendall}}, \bibinfo {author} {\bibfnamefont {R.}~\bibnamefont {Pantone}},
  \bibinfo {author} {\bibfnamefont {K.}~\bibnamefont {Manickavasagam}},
  \bibinfo {author} {\bibfnamefont {Y.}~\bibnamefont {Bengio}},\ and\ \bibinfo
  {author} {\bibfnamefont {B.}~\bibnamefont {Scellier}},\ }\bibfield  {title}
  {\bibinfo {title} {Training {{End-to-End Analog Neural Networks}} with
  {{Equilibrium Propagation}}},\ }\href@noop {} {\bibfield  {journal} {\bibinfo
   {journal} {arXiv:2006.01981 [cs]}\ } (\bibinfo {year} {2020})}\BibitemShut
  {NoStop}%
\bibitem [{\citenamefont {Martin}\ \emph {et~al.}(2021)\citenamefont {Martin},
  \citenamefont {Ernoult}, \citenamefont {Laydevant}, \citenamefont {Li},
  \citenamefont {Querlioz}, \citenamefont {Petrisor},\ and\ \citenamefont
  {Grollier}}]{martin_eqspike_2021}%
  \BibitemOpen
  \bibfield  {author} {\bibinfo {author} {\bibfnamefont {E.}~\bibnamefont
  {Martin}}, \bibinfo {author} {\bibfnamefont {M.}~\bibnamefont {Ernoult}},
  \bibinfo {author} {\bibfnamefont {J.}~\bibnamefont {Laydevant}}, \bibinfo
  {author} {\bibfnamefont {S.}~\bibnamefont {Li}}, \bibinfo {author}
  {\bibfnamefont {D.}~\bibnamefont {Querlioz}}, \bibinfo {author}
  {\bibfnamefont {T.}~\bibnamefont {Petrisor}},\ and\ \bibinfo {author}
  {\bibfnamefont {J.}~\bibnamefont {Grollier}},\ }\bibfield  {title} {\bibinfo
  {title} {{{EqSpike}}: {{Spike-driven}} equilibrium propagation for
  neuromorphic implementations},\ }\href
  {https://doi.org/10.1016/j.isci.2021.102222} {\bibfield  {journal} {\bibinfo
  {journal} {iScience}\ }\textbf {\bibinfo {volume} {24}},\ \bibinfo {pages}
  {102222} (\bibinfo {year} {2021})}\BibitemShut {NoStop}%
\bibitem [{\citenamefont {Stern}\ \emph {et~al.}(2021)\citenamefont {Stern},
  \citenamefont {Hexner}, \citenamefont {Rocks},\ and\ \citenamefont
  {Liu}}]{stern_supervised_2021}%
  \BibitemOpen
  \bibfield  {author} {\bibinfo {author} {\bibfnamefont {M.}~\bibnamefont
  {Stern}}, \bibinfo {author} {\bibfnamefont {D.}~\bibnamefont {Hexner}},
  \bibinfo {author} {\bibfnamefont {J.~W.}\ \bibnamefont {Rocks}},\ and\
  \bibinfo {author} {\bibfnamefont {A.~J.}\ \bibnamefont {Liu}},\ }\bibfield
  {title} {\bibinfo {title} {Supervised {{Learning}} in {{Physical Networks}}:
  {{From Machine Learning}} to {{Learning Machines}}},\ }\href
  {https://doi.org/10.1103/PhysRevX.11.021045} {\bibfield  {journal} {\bibinfo
  {journal} {Physical Review X}\ }\textbf {\bibinfo {volume} {11}},\ \bibinfo
  {pages} {021045} (\bibinfo {year} {2021})}\BibitemShut {NoStop}%
\bibitem [{\citenamefont {Dillavou}\ \emph {et~al.}(2023)\citenamefont
  {Dillavou}, \citenamefont {Beyer}, \citenamefont {Stern}, \citenamefont
  {Miskin}, \citenamefont {Liu},\ and\ \citenamefont
  {Durian}}]{dillavou2023machine}%
  \BibitemOpen
  \bibfield  {author} {\bibinfo {author} {\bibfnamefont {S.}~\bibnamefont
  {Dillavou}}, \bibinfo {author} {\bibfnamefont {B.~D.}\ \bibnamefont {Beyer}},
  \bibinfo {author} {\bibfnamefont {M.}~\bibnamefont {Stern}}, \bibinfo
  {author} {\bibfnamefont {M.~Z.}\ \bibnamefont {Miskin}}, \bibinfo {author}
  {\bibfnamefont {A.~J.}\ \bibnamefont {Liu}},\ and\ \bibinfo {author}
  {\bibfnamefont {D.~J.}\ \bibnamefont {Durian}},\ }\bibfield  {title}
  {\bibinfo {title} {Machine learning without a processor: Emergent learning in
  a nonlinear electronic metamaterial},\ }\href@noop {} {\bibfield  {journal}
  {\bibinfo  {journal} {arXiv preprint arXiv:2311.00537}\ } (\bibinfo {year}
  {2023})}\BibitemShut {NoStop}%
\bibitem [{\citenamefont {Arinze}\ \emph {et~al.}(2023)\citenamefont {Arinze},
  \citenamefont {Stern}, \citenamefont {Nagel},\ and\ \citenamefont
  {Murugan}}]{arinze2023learning}%
  \BibitemOpen
  \bibfield  {author} {\bibinfo {author} {\bibfnamefont {C.}~\bibnamefont
  {Arinze}}, \bibinfo {author} {\bibfnamefont {M.}~\bibnamefont {Stern}},
  \bibinfo {author} {\bibfnamefont {S.~R.}\ \bibnamefont {Nagel}},\ and\
  \bibinfo {author} {\bibfnamefont {A.}~\bibnamefont {Murugan}},\ }\bibfield
  {title} {\bibinfo {title} {Learning to self-fold at a bifurcation},\
  }\href@noop {} {\bibfield  {journal} {\bibinfo  {journal} {Physical Review
  E}\ }\textbf {\bibinfo {volume} {107}},\ \bibinfo {pages} {025001} (\bibinfo
  {year} {2023})}\BibitemShut {NoStop}%
\bibitem [{\citenamefont {Stern}\ \emph
  {et~al.}(2020{\natexlab{a}})\citenamefont {Stern}, \citenamefont {Arinze},
  \citenamefont {Perez}, \citenamefont {Palmer},\ and\ \citenamefont
  {Murugan}}]{stern_supervised_2020}%
  \BibitemOpen
  \bibfield  {author} {\bibinfo {author} {\bibfnamefont {M.}~\bibnamefont
  {Stern}}, \bibinfo {author} {\bibfnamefont {C.}~\bibnamefont {Arinze}},
  \bibinfo {author} {\bibfnamefont {L.}~\bibnamefont {Perez}}, \bibinfo
  {author} {\bibfnamefont {S.~E.}\ \bibnamefont {Palmer}},\ and\ \bibinfo
  {author} {\bibfnamefont {A.}~\bibnamefont {Murugan}},\ }\bibfield  {title}
  {\bibinfo {title} {Supervised learning through physical changes in a
  mechanical system},\ }\href {https://doi.org/10.1073/pnas.2000807117}
  {\bibfield  {journal} {\bibinfo  {journal} {Proceedings of the National
  Academy of Sciences}\ }\textbf {\bibinfo {volume} {117}},\ \bibinfo {pages}
  {14843} (\bibinfo {year} {2020}{\natexlab{a}})}\BibitemShut {NoStop}%
\bibitem [{\citenamefont {Scellier}\ \emph {et~al.}(2023)\citenamefont
  {Scellier}, \citenamefont {Ernoult}, \citenamefont {Kendall},\ and\
  \citenamefont {Kumar}}]{scellier2023energy}%
  \BibitemOpen
  \bibfield  {author} {\bibinfo {author} {\bibfnamefont {B.}~\bibnamefont
  {Scellier}}, \bibinfo {author} {\bibfnamefont {M.}~\bibnamefont {Ernoult}},
  \bibinfo {author} {\bibfnamefont {J.}~\bibnamefont {Kendall}},\ and\ \bibinfo
  {author} {\bibfnamefont {S.}~\bibnamefont {Kumar}},\ }\bibfield  {title}
  {\bibinfo {title} {Energy-based learning algorithms: A comparative study},\
  }in\ \href@noop {} {\emph {\bibinfo {booktitle} {ICML Workshop on Localized
  Learning (LLW)}}}\ (\bibinfo {year} {2023})\BibitemShut {NoStop}%
\bibitem [{\citenamefont {Anisetti}\ \emph {et~al.}(2023)\citenamefont
  {Anisetti}, \citenamefont {Scellier},\ and\ \citenamefont
  {Schwarz}}]{AnisettiPRR2023}%
  \BibitemOpen
  \bibfield  {author} {\bibinfo {author} {\bibfnamefont {V.~R.}\ \bibnamefont
  {Anisetti}}, \bibinfo {author} {\bibfnamefont {B.}~\bibnamefont {Scellier}},\
  and\ \bibinfo {author} {\bibfnamefont {J.~M.}\ \bibnamefont {Schwarz}},\
  }\bibfield  {title} {\bibinfo {title} {Learning by non-interfering feedback
  chemical signaling in physical networks},\ }\href
  {https://doi.org/10.1103/PhysRevResearch.5.023024} {\bibfield  {journal}
  {\bibinfo  {journal} {Phys. Rev. Res.}\ }\textbf {\bibinfo {volume} {5}},\
  \bibinfo {pages} {023024} (\bibinfo {year} {2023})}\BibitemShut {NoStop}%
\bibitem [{\citenamefont {Falk}\ \emph {et~al.}(2023)\citenamefont {Falk},
  \citenamefont {Wu}, \citenamefont {Matthews}, \citenamefont {Sachdeva},
  \citenamefont {Pashine}, \citenamefont {Gardel}, \citenamefont {Nagel},\ and\
  \citenamefont {Murugan}}]{falk2023learning}%
  \BibitemOpen
  \bibfield  {author} {\bibinfo {author} {\bibfnamefont {M.~J.}\ \bibnamefont
  {Falk}}, \bibinfo {author} {\bibfnamefont {J.}~\bibnamefont {Wu}}, \bibinfo
  {author} {\bibfnamefont {A.}~\bibnamefont {Matthews}}, \bibinfo {author}
  {\bibfnamefont {V.}~\bibnamefont {Sachdeva}}, \bibinfo {author}
  {\bibfnamefont {N.}~\bibnamefont {Pashine}}, \bibinfo {author} {\bibfnamefont
  {M.~L.}\ \bibnamefont {Gardel}}, \bibinfo {author} {\bibfnamefont {S.~R.}\
  \bibnamefont {Nagel}},\ and\ \bibinfo {author} {\bibfnamefont
  {A.}~\bibnamefont {Murugan}},\ }\bibfield  {title} {\bibinfo {title}
  {Learning to learn by using nonequilibrium training protocols for adaptable
  materials},\ }\href@noop {} {\bibfield  {journal} {\bibinfo  {journal}
  {Proceedings of the National Academy of Sciences}\ }\textbf {\bibinfo
  {volume} {120}},\ \bibinfo {pages} {e2219558120} (\bibinfo {year}
  {2023})}\BibitemShut {NoStop}%
\bibitem [{\citenamefont {Wycoff}\ \emph {et~al.}(2022)\citenamefont {Wycoff},
  \citenamefont {Dillavou}, \citenamefont {Stern}, \citenamefont {Liu},\ and\
  \citenamefont {Durian}}]{wycoff_desynchronous_2022}%
  \BibitemOpen
  \bibfield  {author} {\bibinfo {author} {\bibfnamefont {J.~F.}\ \bibnamefont
  {Wycoff}}, \bibinfo {author} {\bibfnamefont {S.}~\bibnamefont {Dillavou}},
  \bibinfo {author} {\bibfnamefont {M.}~\bibnamefont {Stern}}, \bibinfo
  {author} {\bibfnamefont {A.~J.}\ \bibnamefont {Liu}},\ and\ \bibinfo {author}
  {\bibfnamefont {D.~J.}\ \bibnamefont {Durian}},\ }\bibfield  {title}
  {\bibinfo {title} {Desynchronous learning in a physics-driven learning
  network},\ }\href {https://doi.org/10.1063/5.0084631} {\bibfield  {journal}
  {\bibinfo  {journal} {The Journal of Chemical Physics}\ }\textbf {\bibinfo
  {volume} {156}},\ \bibinfo {pages} {144903} (\bibinfo {year}
  {2022})}\BibitemShut {NoStop}%
\bibitem [{\citenamefont {Stern}\ \emph {et~al.}(2022)\citenamefont {Stern},
  \citenamefont {Dillavou}, \citenamefont {Miskin}, \citenamefont {Durian},\
  and\ \citenamefont {Liu}}]{stern2022physical}%
  \BibitemOpen
  \bibfield  {author} {\bibinfo {author} {\bibfnamefont {M.}~\bibnamefont
  {Stern}}, \bibinfo {author} {\bibfnamefont {S.}~\bibnamefont {Dillavou}},
  \bibinfo {author} {\bibfnamefont {M.~Z.}\ \bibnamefont {Miskin}}, \bibinfo
  {author} {\bibfnamefont {D.~J.}\ \bibnamefont {Durian}},\ and\ \bibinfo
  {author} {\bibfnamefont {A.~J.}\ \bibnamefont {Liu}},\ }\bibfield  {title}
  {\bibinfo {title} {Physical learning beyond the quasistatic limit},\
  }\href@noop {} {\bibfield  {journal} {\bibinfo  {journal} {Physical Review
  Research}\ }\textbf {\bibinfo {volume} {4}},\ \bibinfo {pages} {L022037}
  (\bibinfo {year} {2022})}\BibitemShut {NoStop}%
\bibitem [{\citenamefont {Patil}\ \emph {et~al.}(2023)\citenamefont {Patil},
  \citenamefont {Ho},\ and\ \citenamefont {Prakash}}]{patil2023self}%
  \BibitemOpen
  \bibfield  {author} {\bibinfo {author} {\bibfnamefont {V.~P.}\ \bibnamefont
  {Patil}}, \bibinfo {author} {\bibfnamefont {I.}~\bibnamefont {Ho}},\ and\
  \bibinfo {author} {\bibfnamefont {M.}~\bibnamefont {Prakash}},\ }\bibfield
  {title} {\bibinfo {title} {Self-learning mechanical circuits},\ }\href@noop
  {} {\bibfield  {journal} {\bibinfo  {journal} {arXiv preprint
  arXiv:2304.08711}\ } (\bibinfo {year} {2023})}\BibitemShut {NoStop}%
\bibitem [{\citenamefont {Hope}(2020)}]{HOPE202067}%
  \BibitemOpen
  \bibfield  {author} {\bibinfo {author} {\bibfnamefont {T.~M.}\ \bibnamefont
  {Hope}},\ }\bibfield  {title} {\bibinfo {title} {Chapter 4 - linear
  regression},\ }in\ \href
  {https://doi.org/https://doi.org/10.1016/B978-0-12-815739-8.00004-3} {\emph
  {\bibinfo {booktitle} {Machine Learning}}},\ \bibinfo {editor} {edited by\
  \bibinfo {editor} {\bibfnamefont {A.}~\bibnamefont {Mechelli}}\ and\ \bibinfo
  {editor} {\bibfnamefont {S.}~\bibnamefont {Vieira}}}\ (\bibinfo  {publisher}
  {Academic Press},\ \bibinfo {year} {2020})\ pp.\ \bibinfo {pages}
  {67--81}\BibitemShut {NoStop}%
\bibitem [{\citenamefont {Lloyd}(1982)}]{lloyd1982least}%
  \BibitemOpen
  \bibfield  {author} {\bibinfo {author} {\bibfnamefont {S.}~\bibnamefont
  {Lloyd}},\ }\bibfield  {title} {\bibinfo {title} {Least squares quantization
  in pcm},\ }\href {https://doi.org/10.1109/TIT.1982.1056489} {\bibfield
  {journal} {\bibinfo  {journal} {IEEE Transactions on Information Theory}\
  }\textbf {\bibinfo {volume} {28}},\ \bibinfo {pages} {129} (\bibinfo {year}
  {1982})}\BibitemShut {NoStop}%
\bibitem [{\citenamefont {Bitzek}\ \emph {et~al.}(2006)\citenamefont {Bitzek},
  \citenamefont {Koskinen}, \citenamefont {G{\"a}hler}, \citenamefont
  {Moseler},\ and\ \citenamefont {Gumbsch}}]{bitzek2006structural}%
  \BibitemOpen
  \bibfield  {author} {\bibinfo {author} {\bibfnamefont {E.}~\bibnamefont
  {Bitzek}}, \bibinfo {author} {\bibfnamefont {P.}~\bibnamefont {Koskinen}},
  \bibinfo {author} {\bibfnamefont {F.}~\bibnamefont {G{\"a}hler}}, \bibinfo
  {author} {\bibfnamefont {M.}~\bibnamefont {Moseler}},\ and\ \bibinfo {author}
  {\bibfnamefont {P.}~\bibnamefont {Gumbsch}},\ }\bibfield  {title} {\bibinfo
  {title} {Structural relaxation made simple},\ }\href@noop {} {\bibfield
  {journal} {\bibinfo  {journal} {Physical review letters}\ }\textbf {\bibinfo
  {volume} {97}},\ \bibinfo {pages} {170201} (\bibinfo {year}
  {2006})}\BibitemShut {NoStop}%
\bibitem [{\citenamefont {Schrauwen}\ \emph {et~al.}(2007)\citenamefont
  {Schrauwen}, \citenamefont {Verstraeten},\ and\ \citenamefont
  {Van~Campenhout}}]{schrauwen2007overview}%
  \BibitemOpen
  \bibfield  {author} {\bibinfo {author} {\bibfnamefont {B.}~\bibnamefont
  {Schrauwen}}, \bibinfo {author} {\bibfnamefont {D.}~\bibnamefont
  {Verstraeten}},\ and\ \bibinfo {author} {\bibfnamefont {J.}~\bibnamefont
  {Van~Campenhout}},\ }\bibfield  {title} {\bibinfo {title} {An overview of
  reservoir computing: theory, applications and implementations},\ }in\
  \href@noop {} {\emph {\bibinfo {booktitle} {Proceedings of the 15th european
  symposium on artificial neural networks. p. 471-482 2007}}}\ (\bibinfo {year}
  {2007})\ pp.\ \bibinfo {pages} {471--482}\BibitemShut {NoStop}%
\bibitem [{\citenamefont {Misra}\ \emph {et~al.}(2023)\citenamefont {Misra},
  \citenamefont {Mitchell}, \citenamefont {Chen},\ and\ \citenamefont
  {Sung}}]{misra2023design}%
  \BibitemOpen
  \bibfield  {author} {\bibinfo {author} {\bibfnamefont {S.}~\bibnamefont
  {Misra}}, \bibinfo {author} {\bibfnamefont {M.}~\bibnamefont {Mitchell}},
  \bibinfo {author} {\bibfnamefont {R.}~\bibnamefont {Chen}},\ and\ \bibinfo
  {author} {\bibfnamefont {C.}~\bibnamefont {Sung}},\ }\bibfield  {title}
  {\bibinfo {title} {Design and control of a tunable-stiffness
  coiled-actuator},\ }\href@noop {} {\bibfield  {journal} {\bibinfo  {journal}
  {IEEE International Conference on Robotics and Automation (ICRA)}\ }
  (\bibinfo {year} {2023})}\BibitemShut {NoStop}%
\bibitem [{\citenamefont {Stern}\ \emph
  {et~al.}(2020{\natexlab{b}})\citenamefont {Stern}, \citenamefont {Pinson},\
  and\ \citenamefont {Murugan}}]{stern_continual_2020}%
  \BibitemOpen
  \bibfield  {author} {\bibinfo {author} {\bibfnamefont {M.}~\bibnamefont
  {Stern}}, \bibinfo {author} {\bibfnamefont {M.~B.}\ \bibnamefont {Pinson}},\
  and\ \bibinfo {author} {\bibfnamefont {A.}~\bibnamefont {Murugan}},\
  }\bibfield  {title} {\bibinfo {title} {Continual {{Learning}} of {{Multiple
  Memories}} in {{Mechanical Networks}}},\ }\href
  {https://doi.org/10.1103/PhysRevX.10.031044} {\bibfield  {journal} {\bibinfo
  {journal} {Physical Review X}\ }\textbf {\bibinfo {volume} {10}},\ \bibinfo
  {pages} {031044} (\bibinfo {year} {2020}{\natexlab{b}})}\BibitemShut
  {NoStop}%
\bibitem [{\citenamefont {Stern}\ \emph
  {et~al.}(2024{\natexlab{b}})\citenamefont {Stern}, \citenamefont {Liu},\ and\
  \citenamefont {Balasubramanian}}]{stern2024physical}%
  \BibitemOpen
  \bibfield  {author} {\bibinfo {author} {\bibfnamefont {M.}~\bibnamefont
  {Stern}}, \bibinfo {author} {\bibfnamefont {A.~J.}\ \bibnamefont {Liu}},\
  and\ \bibinfo {author} {\bibfnamefont {V.}~\bibnamefont {Balasubramanian}},\
  }\bibfield  {title} {\bibinfo {title} {Physical effects of learning},\
  }\href@noop {} {\bibfield  {journal} {\bibinfo  {journal} {Physical Review
  E}\ }\textbf {\bibinfo {volume} {109}},\ \bibinfo {pages} {024311} (\bibinfo
  {year} {2024}{\natexlab{b}})}\BibitemShut {NoStop}%
\end{thebibliography}
\end{document}